%% file: main.tex
\documentclass[sigconf]{acmart}
\usepackage{multirow}
\input{math_commands.tex}

\newcommand{\ie}{\textit{i.e.,~}}
\newcommand{\eg}{\textit{e.g.,~}}

\usepackage[inline]{enumitem}
\setlist[description]{leftmargin=\parindent,labelindent=\parindent, font=\normalfont\itshape}

\usepackage{multirow}
\usepackage{algorithm}
\usepackage{hyperref}
\usepackage{cleveref} % for better referencing; has to be placed before IEEE tran tools, else one sees odd errors

\crefname{section}{\S}{\S} 
\crefname{subsection}{\S}{\S} 

\usepackage[rightcaption]{sidecap}
\usepackage{caption}
\AtBeginDocument{%
  \providecommand\BibTeX{{%
    \normalfont B\kern-0.5em{\scshape i\kern-0.25em b}\kern-0.8em\TeX}}}

\copyrightyear{2024}
\acmYear{2024}
\setcopyright{rightsretained}
\acmConference[RecSys '24]{18th ACM Conference on Recommender Systems}{October 14--18, 2024}{Bari, Italy}
\acmBooktitle{18th ACM Conference on Recommender Systems (RecSys '24), October 14--18, 2024, Bari, Italy}\acmDOI{10.1145/3640457.3688145}
\acmISBN{979-8-4007-0505-2/24/10}

\begin{document}

\title[Pre-trained Sequential Recommender]{A Pre-trained Zero-shot Sequential Recommendation Framework via Popularity Dynamics} 
\newcommand{\name}{{\textsc{\textsc{PrepRec}~}}}

\newtheorem{problem}{Problem}
\newcommand{\subscript}[2]{$#1 _ #2$}

\author{Junting Wang}
\authornote{Both authors contributed equally to this research.}
\email{junting3@illinois.edu}
\affiliation{%
  \institution{University of Illinois at Urbana-Champaign}
  \streetaddress{201 N Goodwin Ave}
  \city{Urbana}
  \state{Illinois}
  \postcode{61801}
  \country{USA}
}

\author{Praneet Rathi}
\authornotemark[1]
\email{prathi3@illinois.edu}
\affiliation{%
  \institution{University of Illinois at Urbana-Champaign}
  \streetaddress{201 N Goodwin Ave}
  \city{Urbana}
  \state{Illinois}
  \postcode{61801}
  \country{USA}
}

\author{Hari Sundaram}
\email{hs1@illinois.edu}
\affiliation{%
  \institution{University of Illinois at Urbana-Champaign}
  \streetaddress{201 N Goodwin Ave}
  \city{Urbana}
  \state{Illinois}
  \postcode{61801}
  \country{USA}
}

\setlength{\abovedisplayskip}{0.1cm}
\setlength{\belowdisplayskip}{0.1cm}
\setlength{\floatsep}{0.1cm}
\setlength{\textfloatsep}{0.1cm}
\setlength{\abovecaptionskip}{0.1cm}
\setlength{\belowcaptionskip}{0.1cm}
\setlength{\dbltextfloatsep}{0.1cm}
\setlength{\intextsep}{0.1cm}

\renewcommand{\shortauthors}{Wang and Rathi, et al.}
\newcommand\barbelow[1]{\stackunder[1.2pt]{$#1$}{\rule{.8ex}{.075ex}}}

\makeatletter
\newcommand{\algmargin}{\the\ALG@thistlm}
\makeatother

\begin{abstract}

\input{abstract.tex}
\end{abstract}
\vspace{-10pt}
%   <ccs2012>
%      <concept>
%          <concept_id>10002951.10003317.10003347.10003350</concept_id>
%          <concept_desc>Information systems~Recommender systems</concept_desc>
%          <concept_significance>500</concept_significance>
%          </concept>
%    </ccs2012>
%   \end{CCSXML}
  
  \vspace{-10pt}

\vspace{-5pt}

\maketitle
\section{INTRODUCTION}
\label{sec:introduction}
\input{introduction.tex}

\vspace{-10pt}
\section{RELATED WORK}
\label{sec:related_work}
\input{related_work.tex}
\vspace{-10pt}
\section{PROBLEM DEFINITION}
\label{sec:problem_formulation}

\input{problem_formulation.tex}
\vspace{-10pt}
\section{PREPREC FRAMEWORK}
\label{sec:methods}
\input{methods.tex}

\vspace{-10pt}
\section{EXPERIMENTS}
\label{sec:experiments}
\input{experiments.tex}
\vspace{-10pt}
\section{CONCLUSION}
\label{sec:conclusion}
\input{conclusion}

\begin{acks}
This work was generously supported by the National Science Foundation (NSF) under grant number 2312561. We also would like to thank the anonymous reviewers for their valuable feedback.
\end{acks}

% \newpage
\bibliographystyle{ACM-Reference-Format}
\bibliography{main,hs}

% \clearpage
% \appendix
% \input{appendix.tex}

\end{document}

%% file: math_commands.tex
\usepackage{amsmath,amsfonts,bm}

\def\eqref#1{equation~\ref{#1}}
\def\1{\bm{1}}

\def\mE{{\bm{E}}}

\def\mM{{\bm{M}}}

\def\mP{{\bm{P}}}

\def\mT{{\bm{T}}}

\def\mW{{\bm{W}}}

\DeclareMathAlphabet{\mathsfit}{\encodingdefault}{\sfdefault}{m}{sl}
\SetMathAlphabet{\mathsfit}{bold}{\encodingdefault}{\sfdefault}{bx}{n}

%% file: abstract.tex
This paper proposes a novel pre-trained framework for zero-shot cross-domain  sequential recommendation without auxiliary information. 
% This paper focuses on the challenging task of zero-shot cross-domain and cross-application sequential recommendation without auxiliary information. 
While using auxiliary information (\eg item descriptions) seems promising for cross-domain transfer, a cross-domain adaptation of sequential recommenders can be challenging when the target domain differs from the source domain---item descriptions are in different languages; metadata modalities (\eg audio, image, and text) differ across source and target domains. If we can learn universal item representations independent of the domain type (\eg groceries, movies), we can achieve zero-shot cross-domain transfer without auxiliary information. Our critical insight is that user interaction sequences highlight shifting user preferences via the popularity dynamics of interacted items.  
We present a pre-trained sequential recommendation framework: \textbf{\textsc{PrepRec}}, which utilizes a novel popularity dynamics-aware transformer architecture. Through extensive experiments on five real-world datasets, we show that \textsc{PrepRec}, without any auxiliary information, can zero-shot adapt to new application domains and achieve competitive performance compared to state-of-the-art sequential recommender models.
In addition, we show that \name complements existing sequential recommenders. With a simple post-hoc interpolation, \name improves the performance of existing sequential recommenders on average by 11.8\% in Recall@10 and 22\% in NDCG@10. We provide an anonymized implementation of \name at \url{https://github.com/CrowdDynamicsLab/preprec}.
% In addition, with a simple post-hoc interpolation, \name can improve the performance of existing sequential recommenders on average by 12.4\% in Recall@10 and 22\% in NDCG@10. We provide an anonymized implementation of \name at \url{https://anonymous.4open.science/r/PrepRec--2F60/}.
\vspace{-5pt}

% Sequential recommenders model sequential users' behavior and are the fundamental building blocks of many online applications, \eg e-commerce, video streaming, and social media. While effective, they learn item embeddings one-on-one mapped to explicit item IDs, which are unique in each domain, making them hard to generalize to new recommendation domains; whereas pre-trained models have shown great success in generalizing to new application domains in peer AI fields. Pioneer works on pre-trained sequential recommenders rely on application-dependent auxiliary information and can only generalize to the related domains within the same application, \eg online retail. In this work, we aim to tackle the most challenging task: zero-shot cross-domain and cross-application sequential recommendation without auxiliary information, to set a baseline for pre-trained sequential recommenders. 

%% file: introduction.tex
% Sequential recommendation, being one of the most studied tasks in recommender systems, aims to predict the next item that a user will interact with, given a user's historical interaction records. Numerous methods have been proposed to tackle this task, from early methods using Markov chains~\cite{rendle2010factorizing, shani2005mdp} to neural network-based methods such as GRU4Rec~\cite{hidasi2015session} and GRU4Rec+~\cite{tan2016improved}. Recently, with the success of attention-based models in natural language processing, some works have explored using Transformers~\cite{attention} and BERT~\cite{devlin2018bert} for sequential recommendation~\cite{sun2019bert4rec, kang2018self, li2020time}. 
% Sequential 

% While these models differ in their architectures, their training objectives and processes are very similar: they model user behavior from their historical interaction sequences and capture user preference as a function of the items in the sequence. Then, they can learn a sequential model for the next item prediction objective. Despite the progress in model architectures, these models are trained on a single domain and learn specific representations for each item~\cite{sun2019bert4rec,kang2018self}. Therefore, they cannot generalize to a new domain or application, due to the limitations of modeling by explicit item IDs. As a result, retraining is required for these models to work in a new domain or even new items in the same domain, which is both time-consuming and computationally expensive. 

Modeling sequential user behavior is critical to the success of online applications such as e-commerce, video streaming, and social media. Despite essential innovations for tackling the sequential recommendation task \cite{rendle2010factorizing,hidasi2015session,tan2016improved, sun2019bert4rec, kang2018self, li2020time, jiao2024rethinking}, these systems have some limitations. Firstly, they must be trained from scratch for each application domain because they learn domain-specific item representations~\cite{sun2019bert4rec,kang2018self, he2024co, he2023robust}, which is resource-consuming and limits model reuse across domains.  Even within a domain, they must be retrained when there is a large influx of users or items to maintain performance. Prior work tackles these limitations by incorporating auxiliary information~\cite{ding2021zero,hou2022towards, hao2021pre}, \eg item descriptions. However, using auxiliary information can be problematic for cross-domain transfer if item descriptions are in different languages (\eg English and Chinese), or if the metadata modalities (\eg, audio, image, and text) differ across domains.

This paper tackles a challenging cross-domain transfer setting where we assume no access to auxiliary information. Thus, we ask: \textit{can we build a pre-trained sequential recommender system capable of cross-domain and cross-application transfer without any auxiliary information?} (\eg using the model trained for online shopping in the US to predict the next movie a user in China will watch). Our work is a performance baseline for cross-domain tasks;  using compatible (\ie same language/modality) auxiliary information across domains, can only improve the performance.
\begin{figure}[t]
    \centering
    \includegraphics[width=\linewidth]{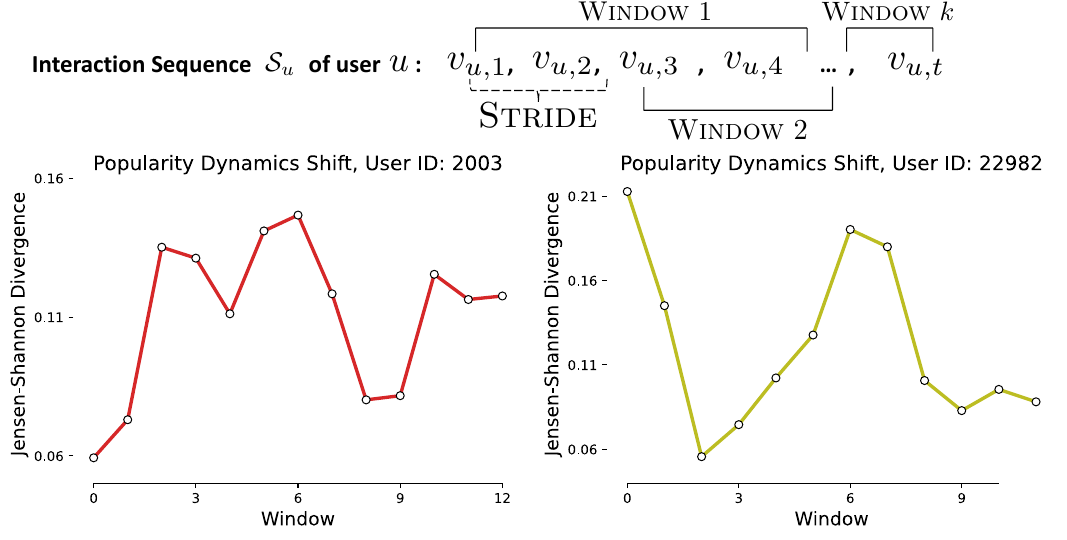}
    \caption{Jensen-Shannon divergence between two consecutive windows ($k$, $k+1$), where we measure the change in popularity of items in the user's sequence. The sampled users are from the Amazon Office dataset. This shows that there exist temporal item popularity shifts in user interaction sequences. We set the window size to 10 and stride to 5.}
    \label{fig:intro}
    \vspace{-5pt}
\end{figure}
% As a result, these models cannot generalize to new domains or even new items in the same domain, due to the limitations of modeling by explicit item IDs. Therefore, while achieving great performance, these models are not practical in real-world applications, where the item space is constantly evolving by the constant influx of new items, which necessitates time-consuming and computationally expensive retraining.

At first glance, developing pre-trained sequential recommenders for cross-domain inference seems impossible. While we see pre-trained language~\cite{bert,gpt3,openai2023gpt4,raffel2020exploring,liu2019roberta, liu2025climb, li2024advances} and vision models~\cite{dosovitskiy2020image,he2016deep,radford2021learning, lin2025moralise} show excellent generalizability across datasets and applications, being able to achieve state-of-the-art performance by just a few fine-tuning steps~\cite{bert,liu2019roberta} or even without any training~\cite{openai2023gpt4,gpt3} (\ie zero-shot transfer), there are essential differences. The representations learned by the pre-trained language model seem universal since the training domain and the application domain (\eg text prediction and generation) share the same language and vocabulary, supporting the effective reuse of the word representations. However, in the cross-domain recommendation, the items are distinct across domains in recommendation datasets (\eg grocery items vs movies). Therefore, forming such generalizable correspondence is nearly impossible if we learn representations for each item \textit{within} each domain. Recent work explores pre-trained models for sequential recommendation~\cite{ding2021zero,hou2022towards, hao2021pre} within the same application (\eg online retail). However, they assume access to metadata of items (\eg item description), which is domain-dependent and is often not generalizable to other domains. These models cannot learn universal representations of items; instead, they bypass the representation learning problem by using additional item-side information.

\textbf{Our Insight:} There exist item popularity shifts in the user's sequence, as indicated in~\Cref{fig:intro}. The item popularity shifts can be explained as temporal shifts in the user's preferences. For example, a user might be interested in buying some common office goods such as pens, papers, and notebooks, but afterward, they might look for other less common office goods such as a whiteboard or a desk. Previous works try to learn users' preference from the past sequence but \textit{ignore the crucial aspect of item popularity dynamics}, which could indicate the user's changing preferences. We know that the marginal distribution of user and item activities are heavy-tailed across datasets, supported by prior work in network science~\cite{Barabasi1999,Barabasi2005} and by experiments in recommender systems~\cite{Salganik2006}. In addition, recent work in recommender systems suggests that the popularity dynamics of items are also crucial for predicting users' behaviors~\cite{ji2020re}.

% Intuitively, how users react to changes in item popularity should be predictable across domains, \ie some users might be more sensitive to changes in item popularity than others. 

\textbf{Present Work:} In this paper, we develop universal, transferable item representations for the zero-shot, cross-domain setting based on the popularity dynamics of items. We explicitly model the popularity dynamics of items and propose a novel pre-trained sequential recommendation framework: \textsc{PrepRec}. We learn universal item representations based on their popularity dynamics instead of their explicit item IDs or auxiliary information. We encode the relative time interval between two consecutive interactions via relative-time encoding and the position of each interaction in the sequence by positional encoding. Using physical time ensures that the predictions are not anti-causal, \ie using the future interactions to predict the present. We propose a popularity dynamics-aware transformer architecture for learning universal sequence representations. We show that it is possible to build a pre-trained sequential recommender system capable of cross-domain and cross-application transfer without any auxiliary information. Our\textbf{ key contributions} are as follows:
% The learned item popularity representations, time-interval and positional encoding are all universal, making it possible to build a pre-trained sequential recommender system capable of cross-domain and cross-application transfer without any auxiliary information. Our key contributions are as follows:
% Contrary to previous sequential recommenders, we do not learn item representations based on their explicit item IDs, rather, we learn to represent items based on their popularity dynamics. In addition, our framework supports zero-shot cross-domain and cross-application transfer without any auxiliary data, making it more practical and generalizable compared to the previous pre-trained sequential recommenders. Our key contributions are as follows: 
\begin{description}[labelsep=3pt, topsep=0pt]
        % \item[Popularity dynamics:] We are the first work to explicitly model the popularity dynamics of items in sequential recommendation. We learn two temporal representations at any time $t$ for each item's popularity: at a coarse and fine-grained level. We represent items' popularity dynamics (\ie representing popularity changes) by concatenating representations over a fixed time interval. We show that the popularity dynamics of items are not only universal across datasets and applications but can significantly boost the performance of sequential recommendation via a simple post-hoc interpolation. 
        \item[Universal item and sequence representations:] We are the first to learn universal item and sequence representations for sequential recommendation \textit{without any auxiliary information} by exploiting item popularity dynamics. In contrast, prior research learns item representations for each item ID or through item auxiliary information. We learn universal item representations by modeling item popularity dynamics of two temporal resolutions: coarse and fine-grained. We learn universal sequence representations using a carefully designed popularity dynamics-aware transformer architecture. These universal item and sequence representations make possible pre-trained sequential recommender systems capable of cross-domain and cross-application transfer without any auxiliary information.
        % \item[Universal item and sequence representations:] We are the first to learn universal item and sequence representations for sequential recommendation by exploiting item popularity dynamics. In contrast, prior research learns item representations for each item ID or through item auxiliary information. We learn two temporal representations using a transformer architecture with optimizations at any time $t$ for each item's popularity: at a coarse and fine-grained level. We represent items' popularity dynamics (\ie representing popularity changes) by concatenating representations over a fixed time interval. Item dynamics are inferrable from the user-item interaction data, and thus, the learned item representations are \textit{transferable} across domains and applications. These item representations make possible pre-trained sequential recommender systems capable of cross-domain and cross-application transfer without any auxiliary information.

    % \item[Popularity dynamics:] We are the first work to explicitly model the popularity dynamics of items in sequential recommendation. We show that the popularity dynamics of items are not only universal across datasets and applications but are crucial for the performance of sequential recommendation. In addition, popularity dynamics can be used to represent items, making it possible to build a pre-trained sequential recommender system capable of cross-domain and cross-application transfer.
    \item[Zero-shot transfer without auxiliary information:] We propose a new challenging setting for pre-trained sequential recommender systems: zero-shot cross-domain and cross-application transfer without any auxiliary information. 
    % We propose a pre-trained sequential recommendation framework: \textsc{PrepRec}, that is capable of zero-shot cross-domain and cross-application transfer without any auxiliary information. 
    In contrast, previous pre-trained sequential recommenders requires overlapping users~\cite{yuan2020parameter},  application-dependent auxiliary information~\cite{ding2021zero,hou2022towards, hao2021pre,hou2023learning, MISSRec}, and are few-shot adapted to related domains within the same application~\cite{hou2022towards, hou2023learning, MISSRec}. Our work establishes a performance baseline for cross-domain sequential recommenders that use compatible (\ie same language/modality) auxiliary information across domains, as such metadata can only improve the performance of cross-domain transfer.
    % In contrast, previous works in sequential recommender systems capable of cross-domain zero-shot rely heavily on application-dependent auxiliary information \cite{ding2021zero,hou2022towards, hao2021pre}. To the best of our knowledge, we are the first to tackle this challenging setting in sequential recommendation.
\end{description}

With extensive experiments, we empirically show that \name has excellent generalizability across domains and applications. Remarkably, had we trained a state-of-the-art model from scratch for the target domain, \textit{instead of zero-shot transfer using} \textsc{PrepRec}, the maximum performance gain over \name would have been only 4\%. In addition, we show that \name is complementary to state-of-the-art sequential recommenders and with a post-hoc interpolation, \name can outperform the state-of-the-art sequential recommender system on average by 11.8\% in Recall@10 and 22\% in NDCG@10. We attribute the improvements to the performance gains over long-tail items, which we show in the qualitative analysis. With this work, we set a baseline for pre-trained sequential recommenders and show that popularity dynamics not only enable us to build a pre-trained sequential recommender system capable of zero-shot transfer but also significantly boost the performance of sequential recommendation.

% With this work, it becomes clear that item popularity dynamics explain much of the prediction ability of the modern neural recommender systems, shedding light on what neural models are learning from data and opening a new direction for improvement in the future. 

% This impressive performance underscores the importance of representing item popularity dynamics for sequential recommendations. 

% Further, this work opens up a new research direction in recommender systems, where we can develop models that use the representations of the item popularity dynamics as a starting point to establish the neural architecture.  

%% file: related_work.tex
\textbf{Sequential Recommendation}: Sequential recommenders model user behavior as a sequence of interactions, and aim to predict the next item that a user will interact with. Early sequential recommenders adopt Markov chains~\cite{rendle2010factorizing, shani2005mdp} and basic neural network architectures~\cite{tang2018personalized, tuan20173d, hidasi2016parallel, hidasi2015session, tan2016improved}.
%Sequential recommenders model user behavior as a sequence of interactions, and aim to predict the next item that a user will interact with. Early sequential recommenders adopt Markov chains and model item-item transition probability~\cite{rendle2010factorizing, shani2005mdp}. Later, with the advance of neural networks, some basic neural architectures have been explored for sequential recommenders, such as convolutional neural networks~\cite{tang2018personalized, tuan20173d}. Recurrent neural networks~\cite{hidasi2016parallel} and its variants, such as GRU4Rec~\cite{hidasi2015session} and GRU4Rec+~\cite{tan2016improved}, have also been widely used for sequential recommendation due to the sequential nature of RNN-based architectures. 
With the success of Transformer~\cite{vaswani2017attention} in modeling sequential data ~\cite{sun2019bert4rec, kang2018self, li2020time} adopt the transformer architecture for sequential recommendation. Additionally, \cite{li2020time} considers the timestamps of each interaction and proposes a time-aware attention mechanism. ~\cite{lv2019sdm, ying2018sequential, tan2021dynamic} separate interaction sequences and categorize them to show the long-term and short-term interests of users. Temporal sequential recommenders~\cite{koren2009collaborative, wu2017recurrent, zhang2014latent} models the change in users' preferences. These works, while achieving state-of-the-art performance, only focus on the regular sequential recommendation and cannot transfer to other domains.

\textbf{Cross-domain Recommendation}: Cross-domain recommendation literature leverages the information-rich domain to improve the recommendation performance on the data-sparse domain~\cite{hu2018conet,li2020ddtcdr,man2017cross}. However, most of these works assume user or item overlap~\cite{zhu2021transfer,catn,hu2018conet,li2020ddtcdr,man2017cross} for effective knowledge transfer. Other cross-domain literature focuses on the cold-start problem~\cite{lu2020meta, zhu2021transfer,  feng2021zero, dropoutnet,wei2021contrastive,du2020learn,liu2021leveraging, melu,dong2020mamo}. In addition, multi-domain recommenders ~\cite{star,ariza2023exploiting} leverage multi-domain data to gain insights into user preferences and item characteristics.

\textbf{Pre-trained Sequential Recommenders}: Recently, pre-trained recommenders have caught the attention of the community. ZESRec~\cite{ding2021zero} is capable of zero-shot sequential recommendations. However, it only works for closely related domains and requires item metadata. PeterRec~\cite{yuan2020parameter} requires overlapping users in both domains. On the other hand, finetuning-based models, \eg MISSRec~\cite{MISSRec}, UnisRec~\cite{hou2022towards}, and VQ-Rec~\cite{hou2023learning}, are not designed for zero-shot sequential recommendation and works within the same application (e-commerce), and they rely on application-dependent auxiliary information. ~\cite{wang2023pretrained} investigates the joint and marginal activity distribution of users and items, but are not suitable for the sequential recommendation task.

To summarize, prior works on sequential recommendation focus on learning high-quality representations for \textit{each} item in the training set and are not generalizable across domains. Pre-trained sequential recommenders are evaluated on closely related domains and platforms and rely heavily on application-dependent auxiliary information of items. 

%% file: problem_formulation.tex
In this section, we formally define the research problems this paper addresses (\ie regular sequential recommendation and zero-shot sequential recommendation) and introduce our notations.

In sequential recommendation, denote $\mM$ as the implicit feedback matrix, $\mathcal{U}$ = $\{u_1, u_2, ..., u_{|\mathcal{U}|}\}$ as the set of users, $\mathcal{V}$ = $\{v_1, v_2, ..., v_{|\mathcal{V}|}\}$ as the set of items.
% , and let $\mathcal{S}_u$ = $\{v_{u,1}, v_{u,2}, ..., v_{u, |\mathcal{S}_u|}\}$ be the sequence of items that user $u$ has interacted with, where $v_{u,i} \in \mathcal{V}$ is the $i$-th item in the sequence. The goal of sequential recommendation is to predict the next item $v_{u,|\mathcal{S}_u|+1}$ that user $u$ will interact with. 
The goal of sequential recommendation is to learn a scoring function, that predicts the next item $v_{u, t}$ given a user $u$ 's interaction history $\mathcal{S}_u = \{v_{u,1}, v_{u,2}, ..., v_{u, t-1}\}$. Note that in this paper, since we model time explicitly, we assume access to the timestamp of each interaction, including the next item interaction. We argue that this is a reasonable assumption since the timestamp of the next interaction is always available in practice. For example, if Alice logs in to Netflix, Netflix will always know when Alice logs in and can predict the next movie for Alice. Formally, we define the scoring function as $\mathcal{F}(v^t|\mathcal{S}_u, \mM)$, where $t$ is the time of the prediction.
%  In this paper, we consider building the sequential recommender system with only implicit feedback, \ie no auxiliary information. 
 
% \textbf{Zero-shot Sequential Recommendation}: In zero-shot sequential recommendation, given an interaction matrix $\mM'$ over $\mathcal{U}'$ and $\mathcal{V'}$, the goal is to produce a scoring function $\mathcal{F}'$ without training on $\mM'$ directly. In other words, the scoring function $\mathcal{F}'$ has to be trained on a different interaction matrix $\mM$. We study the zero-shot sequential recommendation that not only $\mM$ and $\mM'$ are from different domains and platforms, but also both $\mathcal{U} \cap \mathcal{U}' = \emptyset$ and $\mathcal{V} \cap \mathcal{V}' = \emptyset$. Unlike previous pre-trained sequential recommenders that evaluate the model on related domains~\cite{ding2021zero,hou2022towards} or overlapping users~\cite{yuan2020parameter}, we study the most challenging zero-shot sequential recommendation setting, where no auxiliary information is available in all domains.
 \textbf{Zero-shot Sequential Recommendation}: Given two domains $\mM$ and $\mM'$ over $\mathcal{U}, \mathcal{V}$ and $\mathcal{U}', \mathcal{V'}$ respectively, we study the zero-shot recommendation problem in the scenario where the domains are different ($\mM\cap\mM' = \varnothing $), users are disjoint ($\mathcal{U}\cap\mathcal{U'}= \varnothing$), and item sets are unique ($\mathcal{V}\cap\mathcal{V'}= \varnothing$). The goal is to produce a scoring function $\mathcal{F}'$ without training on $\mM'$ directly. In other words, the scoring function $\mathcal{F}'$ has to be trained on a different interaction matrix $\mM$. Furthermore, we assume there is no metadata associated with users or items, which makes the problem particularly challenging but crucial to study. We want to set a baseline for pre-trained sequential recommenders, using metadata can only improve and simply the problem.
 
 % Unlike previous pre-trained sequential recommenders that evaluate the model on related domains~\cite{ding2021zero,hou2022towards} or overlapping users~\cite{yuan2020parameter}, we study the most challenging zero-shot sequential recommendation setting, where no auxiliary information is available in all domains.

% Compared to previous works on zero-shot sequential recommendation, which assume access to side information, we pose a more challenging problem setting. Our experimental setting is more universal and practical, and requires the model to learn the truly universal properties of the data. In other words, the zero-shot sequential recommendation problem we study is the most challenging setting, where no auxiliary information is available in both domains.

%% file: methods.tex
\begin{figure*}[t]
    \centering
    \includegraphics[width=0.8\textwidth]{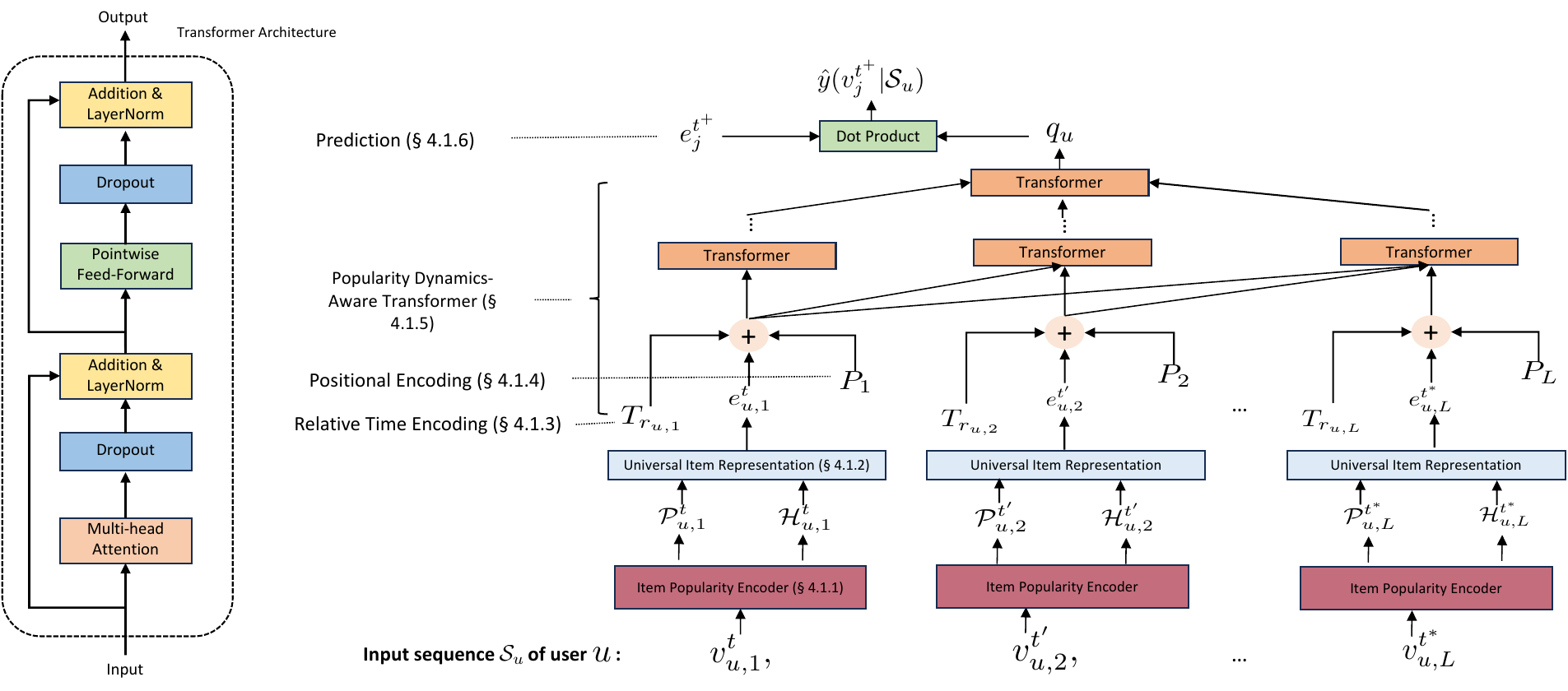}
    \caption{Model Architecture of \name}
    \label{fig:model_architecture}
\end{figure*}

We first introduce the model architecture of \name (\cref{sec:model_architecture}) then the training procedure (\cref{sec:training_procedure}). Finally, we formally define the zero-shot inference process (\cref{sec:zero_shot_inference}).

\vspace{-5pt}
\subsection{Model Architecture}
\label{sec:model_architecture}
\vspace{-2pt}

The first step of building a pre-trained sequential recommender is to learn universal item representations. Our solution is to exploit the item popularity statistics to learn universal item representations. We learn to represent items at a given timestamp through the changes in their popularity histories over different periods, \ie popularity dynamics. We propose a popularity dynamics-aware Transformer architecture that obtains the representation of users' behavior sequences through item popularity dynamics. 
% To learn the universal item representations, we propose a novel item representation encoder that learns to represent items through the changes in their popularity histories over different periods (popularity dynamics).
% Previous works in sequential and pre-trained sequential recommender systems learn to represent items either through learnable embeddings specific to their item IDs or through their textual information. To learn the . In this work, we propose a novel item representation encoder that learns to represent items through the changes in their popularity histories over different periods (popularity dynamics).
\vspace{-5pt}
\subsubsection{Item Popularity Encoder}
\label{sec:item_popularity_encoding}
We learn to represent items based on their popularity dynamics, \ie changes in their popularity histories. Intuitively, popularity can be calculated by two horizons: long-term and short-term. Long-term horizons reflect the overall popularity of items, whereas short-term horizons should capture the recent trends in the domain. For example, the long-term popularity of a winter coat measures how popular is the coat in general, while its short-term popularity reflects more temporal changes, \eg season, weather conditions, and fashion trends. Therefore, consider an item $v_j$ that has interaction at time $t$, denoted as $v_j^t$, we define two popularity representations for $v_j^t$: popularity $\mathbf{p}_j^t \in \mathbb{R}^{k}$ over a coarse period (\eg month) and popularity $\mathbf{h}_j^t \in \mathbb{R}^{k}$ over a fine period (\eg week).

To calculate $\mathbf{p}_j^t$ and $\mathbf{h}_j^t$, we first calculate the popularities of $v_j^t$ over the two horizons, denoted as $a_j^t\in \mathbb{R}^+$ (coarse period number of interactions) and $b_j^t\in \mathbb{R}^+$ (fine period number of interactions). Specifically, we calculate them as: 
\begin{equation}
        a_j^t = \sum_{m=1}^{t} \gamma^{t-m} c_a(v_j^m), \quad b_j^t = c_b(v_j^t)
    \label{eqn:popularity_gamma}
\end{equation}
where $\gamma\in \mathbb{R}^+$ is a pre-defined discount factor and $c_a(v_j^m)$ is the number of interactions of $v_j$ over a coarse time period $m$. Similarly, $c_b(v_j^t)$ denotes the number of interactions of $v_j$ over a fine period $t$. We do not impose the discounting factor when computing $b_j^t$ since we want it to capture the current popularity information, whereas $a_j^t$ captures the cumulative popularity of an item over a longer horizon. 

%The popularity of an item is a real value, which is not comparable across items from different domains. 

To make item popularity comparable across domains, we calculate the percentiles of $a_j^t$ and $b_j^t$ relative to their corresponding coarser and finer popularity distributions over all items at time $t$, denoted as $P(a_j^t) \in \mathbb{R}^+$ and $P({b_j^t})\in \mathbb{R}^+$, respectively.

We now encode the popularity percentiles $P(a_j^t)$ and $P({b_j^t})$ into $k$ dimensional vector representations $\mathbf{p}_j^t$ and $\mathbf{h}_j^t$ respectively. Denote the popularity encoder as $E_p: \mathbb{R}^+ \rightarrow \mathbb{R}^k$, which takes in a percentile value. Suppose given the popularity percentile $P(a_j^t) \in \mathbb{R}^+$ over a coarse time period $t$, the coarse level popularity vector representation $\mathbf{p}_j^t \in \mathbb{R}^k$ is computed as follows:
\begin{align*}
        % E_p(P) &= [\mathbf{p}_{j,1}^t, \mathbf{p}_{j,2}^t, \ldots, \mathbf{p}_{j,11}^t] \\
        % \mathbf{p}_j^t &= E_p(P(a_j^t)) \\
        % (\mathbf{p}_{j}^t)_i &= \begin{cases}
        % 1 - \{\frac{P}{10}\}, & \text{if } i = \lfloor \frac{P}{10} \rfloor \\
        % \{\frac{P}{10}\}, & \text{if } i = \lfloor \frac{P}{10} \rfloor + 1 \\
        % 0, & \text{otherwise}
        \mathbf{p}_j^t &= E_p(P(a_j^t)) \\
        (\mathbf{p}_{j}^t)_i &= \begin{cases}
        1 - \{\frac{P}{k-1}\}, & \text{if } i = \lfloor \frac{P}{k-1} \rfloor \\
        \{\frac{P}{k-1}\}, & \text{if } i = \lfloor \frac{P}{k-1} \rfloor + 1 \\
        0, & \text{otherwise}
        \end{cases}
    \label{eqn:popularity}
\end{align*}
where $\lfloor \cdot \rfloor$ denotes the floor, $\{\cdot\}$ denotes the fractional part of a number, and $(\mathbf{p}_{j}^t)_i$ denotes the $i$-th index of $\mathbf{p}_j^t$. For example, if $k=11$, $E_p(40.1) = [0, 0, 0, 0, 0.99, 0.01, 0, 0, 0, 0, 0]$. 
The interpretation of this would be considering the $10$ deciles for $i \in \{0, 1, \ldots, 9, 10\}$ as basis vectors, and this 
popularity encoding as a linear combination
of the nearest (in percentile space) two basis vectors. The fine level popularity vector is calculated identically, \ie $\mathbf{h}_j^t = E_p(P(b_j^t))$. In this example, we've fixed the vector representation size to be 11, but this approach is fully generalizable to other sizes and would just 
require changing the multipliers in the encoding function. We also experimented with sinuoisal encodings of the same size, but found that the linear encoding empirically performed better.

% Then, we calculate $a_j^t$ and $b_j^t$'s percentiles relative to their corresponding coarser and finer popularity distributions over all items at time $t$, denoted as $P(a_j^t) \in \mathbb{R}$ and $P({b_j^t})\in \mathbb{R}$, respectively. Finally, we encode $P(a_j^t)$ and $P({b_j^t})$ into $k$ dimension vector representations $\mathbf{p}_j^t$ and $\mathbf{h}_j^t$ respectively. The encoding details can be found in~\Cref{sec:popularity_encoding}.

% $a_j^t = \sum_{m=1}^{t} (\gamma_a)^{t-m} c(v_j^m)$, where $\gamma_a$ is the discount factor and $c(v_j^m)$ is the number of interactions of $v_j$ over a coarser time period $m$. Similarly, we define the popularity of $v_j^t$ over a finer time period as $b_j^t = \sum_{n=1}^{t} (\gamma_b)^{t-n} c(v_j^n)$, where $\gamma_b$ is the discount factor and $c(v_j^n)$ is the number of interactions of $v_j$ over a finer time period $n$. Then, we calculate $a_j^t$ and $b_j^t$'s percentiles relative to their corresponding coarser and finer popularity distributions over all items at time $t$, and denote this by $P(a_j^t)$ and $P({b_j^t})$, respectively. We then encode $P(a_j^t)$ and $P({b_j^t})$ into $k$ dimension vector representations $\mathbf{p}_j^t$ and $\mathbf{h}_j^t$ respectively. The encoding details can be found in~\Cref{sec:popularity_encoding}.

\vspace{-5pt}
\subsubsection{Universal Item Representation}
\label{sec:item_trajectory_encoder}
We now define the popularity dynamics of $v_j$ at time $t$ over the coarse period (long-term horizon) to be $\mathcal{P}_j^t= \{\mathbf{p} _j^{1}, \mathbf{p}_j^{2}, ..., \mathbf{p}_j^{t-1}\}$, and over the fine period (short-term horizon) as $\mathcal{H}_j^t = \{\mathbf{h}_j^{1}, \mathbf{h}_j^{2}, ..., \mathbf{h}_j^{t-1}\}$. We use $t-1$ to constrain access to future interactions and prevent information leakage, \ie we do not have access to the popularity statistics of $v_j$ at time $t$ if we are at time $t$. For example, say an interaction happens on the second Wednesday in February, we consider the coarser and finer time period up until the end of January and the end of the first week in February respectively.  To limit computation, we constrain window sizes $m, n$ for $\mathcal{P}$ and $\mathcal{H}$ respectively. Formally, the coarse popularity dynamics of $v_j$ at time $t$ is $\mathcal{P}_j^t = \{\mathbf{p} _j^{t-m}, \mathbf{p}_j^{t-m+1}, ..., \mathbf{p}_j^{t-1}\}$, and the fine popularity dynamics of $v_j$ at time $t$ is $\mathcal{H}_j^t = \{\mathbf{h}_j^{t-n}, \mathbf{h}_j^{t-n+1}, ..., \mathbf{h}_j^{t-1}\}$.
% praneet SUGGESTION: We now define the popularity trajectory of $v_j$ over a coarser time period to be $\{\mathbf{p} _j^{1}, \mathbf{p}_j^{2}, ..., \mathbf{p}_j^t\}$, and over a finer time period as $\{\mathbf{h}_j^{1}, \mathbf{h}_j^{2}, ..., \mathbf{h}_j^t\}$. To limit computation, we consider window sizes $m, n$ for the two trajectories, and formally define the coarse popularity dynamics of $v_j$ as $\mathcal{P}_j^t = \{\mathbf{p} _j^{t-m+1}, \mathbf{p}_j^{t-m+2}, ..., \mathbf{p}_j^t\}$, and the fine popularity dynamics of $v_j$ as $\mathcal{H}_j^t = \{\mathbf{h}_j^{t-n+1}, \mathbf{h}_j^{t-n+2}, ..., \mathbf{h}_j^t\}$.

% We denote the popularity trajectory of $v_j$ over a coarser time period as $\mathcal{P}_j^t = \{p_j^{t-m}, p_j^{t-m+1}, ..., p_j^t\}$ and the popularity trajectory of $v_j$ over a finer time period as $\mathcal{H}_j^t = \{h_j^{t-n}, h_j^{t-n+1}, ..., h_j^t\}$, where $m, n$ are the tunable parameter of indicating the window size we track the two sets of trajectories.

Finally, we compute the embedding of item $v_j$ at time $t$ via the universal item representation encoder, defined as a function $\mathcal{E}(\mathcal{P}_j^t, \mathcal{H}_j^t)$ that learns to encode the popularity dynamics $\mathcal{P}_j^t$ and $\mathcal{H}_j^t$ into a $d$ dimension vector representation $\mathbf{e}_j^t$. Specifically, we have:
\begin{equation} \label{eqn: item_trajectory_encoder}
        \mathbf{e}_j^t = \mathcal{E}(\mathcal{P}_j^t, \mathcal{H}_j^t) = \mW_p[(\|_{i=t-m}^{t-1} \mathbf{p}_j^i) \| (\|_{i=t-n}^{t-1} \mathbf{h}_j^i)]
    % \begin{split}
    %     \mathbf{e}_j^t &= \mathcal{E}(\mathcal{P}_j^t, \mathcal{H}_j^t) = \mW_p[(\concat_{i=t-m}^{t-1} \mathbf{p}_j^i) \oplus (\concat_{i=t-n}^{t-1} \mathbf{h}_j^i)]\\
    %     \concat_{i=t-m}^{t-1} \mathbf{p}_j^i &= \mathbf{p} _j^{t-m}\oplus \mathbf{p}_j^{t-m+1}\oplus ...\oplus \mathbf{p}_j^{t-1}\\
    %     \concat_{i=t-n}^{t-1} \mathbf{h}_j^i &= \mathbf{h}_j^{t-n} \oplus \mathbf{h}_j^{t-n+1}\oplus ...\oplus \mathbf{h}_j^{t-1}
    %             % &= \mW_p(\mathbf{p} _j^{t-m}\oplus \mathbf{p}_j^{t-m+1}\oplus ...\oplus \mathbf{p}_j^{t-1} \oplus \mathbf{h}_j^{t-n} \oplus \mathbf{h}_j^{t-n+1}\oplus ...\oplus \mathbf{h}_j^{t-1}) = \mW_p(\concat_{i=t-m}^{t-1} \mathbf{p}_j^i \oplus \concat_{i=t-n}^{t-1} \mathbf{h}_j^i)
    % \end{split}
\end{equation}
% \begin{equation}
%     \begin{split}
%         \concat_{i=t-m}^{t-1} \mathbf{p}_j^i &\coloneq \mathbf{p} _j^{t-m} \concat \mathbf{p}_j^{t-m+1}\concat...\concat \mathbf{p}_j^{t-1}\\
%         \concat_{i=t-n}^{t-1} \mathbf{h}_j^i &\coloneq \mathbf{h}_j^{t-n} \concat \mathbf{h}_j^{t-n+1}\concat ...\concat \mathbf{h}_j^{t-1}
%     \end{split}
%     \label{eqn:concat}
% \end{equation}
where $\|$ denotes the concatenation operation, and $\mW_p \in \mathbb{R}^{d \times k(m + n)}$ is a learnable weight matrix. 

In addition, we define $\|_{i=t-m}^{t-1} \mathbf{p}_j^i \coloneq \mathbf{p} _j^{t-m} \| \mathbf{p}_j^{t-m+1}\|...\| \mathbf{p}_j^{t-1}$ and $\|_{i=t-n}^{t-1} \mathbf{h}_j^i \coloneq \mathbf{h}_j^{t-n} \| \mathbf{h}_j^{t-n+1}\| ...\| \mathbf{h}_j^{t-1}$. The item popularity dynamics encoder can effectively capture the popularity change of items over different time periods. Most importantly, it does not take explicit item IDs or auxiliary information as input to compute the item embeddings. Instead, it learns to represent items through their popularity dynamics, which is universal across domains and applications.

\subsubsection{Relative Time Interval}
\label{sec:time_interval_encoder}
We also consider the time interval between two consecutive interactions when modeling sequences. Differences in time intervals might indicate differences in the users' behaviors. While previous works explore absolute time intervals~\cite{li2020time}, different domains exhibit diverse time scales, thus making modeling absolute time intervals ungeneralizable. Therefore, we propose to encode relative time intervals into modeling sequences. Given an interaction sequence $\mathcal{S}_u = \{v_{u,1}, v_{u,2}, ..., v_{u,L}\}$ of user $u$, we define the time interval between $v_{u,j}$ and $v_{u,j+1}$ as $t_{u,j} = t(v_{u,j+1}) - t(v_{u,j})$, where $t(v_{u,j})$ is the time that user $u$ interacts with item $v_{u,j}$. We then rank the time intervals of user $u$. Define the rank of relative time interval of $t_{u,j}$ as $r_{u,j} = \text{rank}(t_{u,j})$. The relative time interval encoding of interval $t_{u,j}$ is then defined as $\mT_{r_{u, j}} \in \mathbb{R}^d$, where $\mT \in \mathbb{R}^{L \times d}$, following the same setup in~\cite{vaswani2017attention}, is a fixed sinusoidal encoding matrix defined as:
\begin{equation} \label{eqn: time_interval_encoding}
    % \begin{split}
        \mT_{i, 2j} = \sin(\frac{i}{L^{2j/d}}), \quad\mT_{i, 2j+1} = \cos(\frac{i}{L^{2j/d}})
   %  \end{split}
\end{equation}

We also tried a learnable time interval encoding, but it yielded worse performance. We hypothesize that the sinusoidal encoding is more generalizable across domains and the learnable encoding is more prone to overfitting.

%We then define the relative time interval trajectory of user $u$ as $\mathcal{T}_u = \{r_{u,1}, r_{u,2}, ..., r_{u,L}\}$. We then define the relative time interval encoder as a function $\mathcal{E}_r(\mathcal{T}_u)$ that learns to encode the relative time interval trajectory $\mathcal{T}_u$ into a fixed-length vector representation $\mathbf{e}_u^r$.
% In addition to the item trajectory encoder, we also propose a time interval encoder that learns to encode the time intervals between consecutive interactions of a user. Formally, given an interaction sequence $\mathcal{S}_u = \{v_{u,1}, v_{u,2}, ..., v_{u,|\mathcal{S}_u|}\}$ of user $u$, we define the time interval between $v_{u,i}$ and $v_{u,i+1}$ as $t_{u,i} = t(v_{u,i+1}) - t(v_{u,i})$, where $t(v_{u,i})$ is the time that user $u$ interacts with item $v_{u,i}$. We then define the time interval trajectory of user $u$ as $\mathcal{T}_u = \{t_{u,1}, t_{u,2}, ..., t_{u,|\mathcal{S}_u|-1}\}$. We then define the time interval encoder as a function $\mathcal{E}_t(\mathcal{T}_u)$ that learns to encode the time interval trajectory $\mathcal{T}_u$ into a fixed-length vector representation $\mathbf{e}_u^t$.

\subsubsection{Positional Encoding}
\label{sec:positional_encoding}
As we will see in~\Cref{sec:self_attention}, the self-attention mechanism does not take the positions of the items into account. Therefore, following~\cite{vaswani2017attention}, we also inject a fixed positional encoding for each position in a user's sequence. Denote the positional embedding of a position $l$ as $\mP_l \in \mathbb{R}^d$, where $\mP \in \mathbb{R}^{L \times d}$. We compute $\mP$ using the same formula in~\Cref{eqn: time_interval_encoding}. Again, we also tried a learnable positional encoding as presented in~\cite{kang2018self,sun2019bert4rec}, but it yielded worse results.
% , we define a set of learnable positional embedding to encode the position of each item in the sequence. Formally, given an interaction sequence $\mathcal{S}_u = \{v_{u,1}, v_{u,2}, ..., v_{u,L}\}$ of user $u$, we define the positional embedding of $v_{u,i}$ as $\mathbf{e}_{u,i}^p$. We then define the positional embedding encoder as a function $\mathcal{E}_p(\mathcal{S}_u)$ that learns to encode the positional embedding $\mathbf{e}_{u,i}^p$ into a fixed-length vector representation $\mathbf{e}_u^p$.

\subsubsection{Popularity Dynamics-Aware Transformer}
\label{sec:self_attention}
We follow previous works in sequential recommendation~\cite{kang2018self,sun2019bert4rec,li2020time} and propose an extension to the self-attention mechanism by incorporating universal item representations (\Cref{sec:item_trajectory_encoder}), relative time intervals(\Cref{sec:time_interval_encoder}), and positional encoding (\Cref{sec:positional_encoding}). 
% We propose a novel popularity dynamics-aware self-attention mechanism that learns to capture the temporal changes in item trends and relative time intervals between consecutive interactions of a user. 

Firstly, we transform the user sequence $\{v_{u,1}, v_{u,2}, ..., v_{u, |\mathcal{S}_u|}\}$ for each user $u$ into a fixed-length sequence $\mathcal{S}_u$ = $\{v_{u,1}, v_{u,2}, ..., v_{u, L}\}$ via truncating the oldest interactions or padding, where $L$ is a pre-defined hyper-parameter controlling the maximum length of the sequence. Given a user sequence $\mathcal{S}_u = \{v_{u,1}, v_{u,2}, ..., v_{u,L}\}$, we compute its input matrix $\mE_{u}$ as:  
\begin{equation}
    \mE_u = 
\left[
    \begin{aligned}
         & \mathbf{e}^t_{u, 1} + \mT_{r_{u, 1}} + \mP_1\\
         &\mathbf{e}^{t'}_{u, 2} + \mT_{r_{u, 2}} + \mP_2 \\
         & \vdots \\
        & \mathbf{e}^{t*}_{u, L} + \mT_{r_{u, L}} + \mP_L
    \end{aligned}
    \right]
    \label{eqn:input_embedding}
\end{equation}  
$\mathbf{e}^t_{u, 1}, \mathbf{e}^{t'}_{u, 2},...,\mathbf{e}^{t*}_{u, L}$ is computed from~\Cref{eqn: item_trajectory_encoder}, $\mT_{r_{u, 1}}, \mT_{r_{u, 2}}, ..., \mT_{r_{u, L}}$ and $\mP_1, \mP_2, ..., \mP_L$ are computed following the procedure in~\Cref{sec:time_interval_encoder} and~\Cref{sec:positional_encoding} respectively.

\textbf{Multi-Head Self-Attention}. We adopt a widely used multi-head self-attention mechanism~\cite{vaswani2017attention}, \ie Transformers. Specifically, it consists of multiple multi-head self-attention layers (denoted as MHAttn($\cdot$)), and point-wise feed-forward networks (FFN($\cdot$)). The multi-head self-attention mechanism is defined as:
\begin{equation} \label{eqn: multi_head_attention}
    \begin{split}
        \mathbf{z}_u &= \text{MHAttn}(\mE_u)\\
        \text{MHAttn}(\mE_u) &= \text{Concat}(\text{head}_1, ..., \text{head}_h) \mW^O \\
        %  \text{MHAttn}(\mE_u) &= \concat_{i=1}^h head_i \mW^O \\
        \text{head}_i &= \text{Attn}(\mE_u\mW_i^Q, \mE_u\mW_i^K, \mE_u\mW_i^V)
    \end{split}
\end{equation}
where $\mE_u$ is the input matrix computed from~\Cref{eqn:input_embedding}, $h$ is a tunable hyper-parameter indicating the number of attention heads, $\mW_i^Q, \mW_i^K, \mW_i^V \in \mathbb{R}^{d \times d/h}$ are the learnable weight matrices, and $\mathbf{W}^O \in \mathbb{R}^{d \times d}$ is also a learnable weight matrix. Attn is the attention function and is formally defined as:
\begin{equation} \label{eqn: attention}
    \begin{split}
        \text{Attn}(\mathbf{Q}, \mathbf{K}, \mathbf{V}) &= \text{softmax}(\frac{\mathbf{Q}\mathbf{K}^T}{\sqrt{d/h}})\mathbf{V}
    \end{split}
\end{equation}
The scale factor $\sqrt{d/h}$ is used to avoid large values of the inner product, which can lead to numerical instability.

\textbf{Causality}: In sequential recommendation, the prediction of the $L+1$ item should only depend on the first $L$ items that the user has interacted with in the past. However, the $L$-th output of the multi-head self-attention layer contains all the input information. Therefore, as in \cite{li2020time,kang2018self}, we do not let the model attend to the future items by forbidding links between $Q_i$ and $K_j$ ($j > i$) in the attention function. 

\textbf{Point-Wise Feed-Forward Network}: To add nonlinearity and interactions between different embedding dimensions, we follow previous works in sequential recommendation \cite{sun2019bert4rec,kang2018self,li2020time} and apply the same point-wise feed-forward network to the output of each multi-head self-attention layer. Formally, suppose the output of the multi-head self-attention layer is $\textbf{z}_u$, the point-wise feed-forward network is defined as:
\begin{equation} \label{eqn: feed_forward}
    \begin{split}
        \text{FFN}(\textbf{z}_u) &= \text{ReLU}(\textbf{z}_u\mW_1 + \mathbf{b}_1)\mW_2 + \mathbf{b}_2
    \end{split}
\end{equation}
where $\mW_1 \in \mathbb{R}^{d \times d}$ and $\mW_2 \in \mathbb{R}^{d \times d}$ are learnable weight matrices, and $\mathbf{b}_1 \in \mathbb{R}^{d}$ and $\mathbf{b}_2 \in \mathbb{R}^{d}$ are learnable bias vectors.

\textbf{Stacking Layers}: As shown in previous works~\cite{kang2018self}, stacking multiple multi-head self-attention layers and point-wise feed-forward networks can potentially lead to overfitting and instability during the training. Therefore, we follow previous works~\cite{kang2018self,sun2019bert4rec,li2020time} and apply layer normalization~\cite{ba2016layer} and residual connections to each multi-head self-attention layer and point-wise feed-forward network. Formally, we have:
\begin{equation}
        g(\mathbf{x}) = \mathbf{x} + \text{Dropout}(g(\text{LayerNorm}(\mathbf{x})))
        \label{eqn:residual_connection}
\end{equation}
$g(\mathbf{x})$ is either the multi-head self-attention layer or the point-wise feed-forward network. Therefore, for every multi-head self-attention layer and point-wise feed-forward network, we first apply layer normalization to the input, then apply the multi-head self-attention layer or point-wise feed-forward network, and finally apply dropout and add the input $\mathbf{x}$ to the layer output. The LayerNorm function is defined as:
\begin{equation}
    \text{LayerNorm}(\mathbf{x}) = \alpha \odot \frac{\mathbf{x} - \mu}{\sqrt{\sigma^2 + \epsilon}} + \mathbf{\beta}
    \label{eqn:layer_norm}
\end{equation}
where $\odot$ denotes the element-wise product, $\mu$ and $\sigma$ are the mean and standard deviation of $\mathbf{x}$, $\alpha$ and $\mathbf{\beta}$ are learnable parameters, and $\epsilon$ is a small constant to avoid numerical instability.
% Given a user sequence $\mathcal{S}_u = \{v_{u,1}, v_{u,2}, ..., v_{u,L}\}$, for each item $v_{u,i}$, we compute the item popularity dynamics embedding $\mathbf{e}_{u,i}^t$ using the item trajectory encoder (\cref{sec:item_trajectory_encoder}), the time interval embedding $\mathbf{e}_{u,i}^r$ using the time interval encoder (\cref{sec:time_interval_encoder}), and the positional embedding $\mathbf{e}_{u,i}^p$ of each item $v_{u,i}$ using the positional embedding encoder (\cref{sec:positional_encoding}). The input matrix to the self-attention is then defined as $\mE^i = \{\mathbf{e}_{u,1}^t \oplus \mathbf{e}_{u,1}^r \oplus \mathbf{e}_{u,1}^p, \mathbf{e}_{u,2}^t \oplus \mathbf{e}_{u,2}^r \oplus \mathbf{e}_{u,2}^p, ..., \mathbf{e}_{u,L}^t \oplus \mathbf{e}_{u,L}^r \oplus \mathbf{e}_{u,L}^p\}$. We then apply the multi-head self-attention mechanism (\cref{eqn: multi_head_attention}) to the input matrix $\mE^i$ to obtain the output matrix $\mO^i$. We then apply the point-wise feed-forward network to the output matrix $\mO^i$ to obtain the final output matrix $\mH^i$. Formally, we have:

\subsubsection{Prediction}
\label{sec:prediction}
Given a sequence $\mathcal{S}_u$ of user $u$ as input, we denote $\mathbf{q}_u$ as the output of the popularity dynamics-aware transformer. Suppose at time $t^+$, we want to predict the likelihood of $v_j$ being the next item in the sequence, we first compute the item representation $\mathbf{e}_j^{t^+}$ from~\Cref{sec:item_trajectory_encoder}. Then, we predict the score as the inner product of $\mathbf{q}_u$ and $\mathbf{e}_j^{t^+}$, formally:
\begin{equation}
    \hat{y}(v^t_j|\mathcal{S}_u) = <\mathbf{q}_u, \mathbf{e}_j^{t^+}>
    \label{eqn:prediction}
\end{equation}
Note that there is no information leakage in the prediction process, \ie we do not assume access to the popularity statistics of $v_j$ at time $t^+$ if we are at time $t^+$ (\Cref{sec:item_trajectory_encoder}).
% the reason we use the popularity dynamics embedding of $v_j$ at time $t-1$ is to prevent information leakage, \ie we do not have access to the popularity dynamics of $v_j$ at time $t$ if we are at time $t$.

\subsection{Training Procedure}
\label{sec:training_procedure}
Now we present how to train the \name model. Similar to~\cite{kang2018self}, we adopt the binary cross entropy loss as the objective function, formally:
\begin{equation}
    \begin{split}
        \mathcal{L} = -\sum_{\mathcal{S}_u \in \mathcal{S}} \sum_{z\in [1,2,...,L-1] } &[\log \sigma(\hat{y}(v^t_{z+1}|\mathcal{S}_{u, :z})) \\
         & + \sum_{j' \notin \mathcal{S}_u}\log \sigma(1 -\hat{y}(v^t_{j'}|\mathcal{S}_{u, :z}))]\\
    \end{split}
    \label{eqn:loss}
\end{equation}
where $\mathcal{S}_{u, :z} \coloneq \{v_{u, 1}, v_{u, 2},...,v_{u,z}\}$. $v^t_{z+1}$ represents the $z+1$-th item in the sequence that happened at time $t$. We use Adam~\cite{kingma2014adam} as the optimizer and train the model end-to-end. Note that compared to previous sequential recommenders, we do not have any parameters modeling item IDs. Essentially, we are only optimizing $\mW_p$ and parameters related to the multi-head self-attention mechanism.
\newcommand*{\factor}{0.04}
\begin{table}[t]
    \centering
    \small
    \begin{tabular}{@{}llllll@{}}
    \toprule
    \textsc{Dataset}  &  {\textsc{\#users}} & {\textsc{\#items}}& {\textsc{\#actions}}& {\textsc{avg length}}& {\textsc{density}}\\ 
    \midrule
    Office  &101,133&27,500&0.74M&7.3&0.03\%\\
    Tool  &240,464&73,153&1.96M&8.1&0.01\%\\ 
    Movie  &70,404&40,210&11.55M&164.2&0.41\%\\
    Music  &20,539&10,121&0.66M&32.2&0.32\%\\
    Epinions  &30,989&20,382&0.54M&17.5&0.09\%\\
    \bottomrule
    \end{tabular}
    \caption{Dataset statistics}
    \label{tab:dataset_statistics}
    \vspace{-5pt}
    \end{table}

        % \begin{table}[b]
        %     \centering
        %         \small
        %     \begin{tabular}{p{0.12\linewidth}p{0.1\linewidth}p{0.1\linewidth}p{0.1\linewidth}p{0.1\linewidth}p{0.1\linewidth}
        %     p{0.1\linewidth}}
        %     \toprule
        %     \textsc{Dataset}& {\textsc{Office}} & {\textsc{Tool}} & {\textsc{Movie}} & {\textsc{Music}} & {\textsc{Epinions}}\\ 
        %     \midrule
        %     SasRec &1.331M&3.581M&2.044M&0.542M&1.054M\\
        %     BERT4Rec &2.687M&7.233M&4.126M&1.094M&2.127M\\
        %     TiSasRec &1.367M&3.617M&2.127M&0.578M&1.09M\\
        %     \name &0.045M&0.045M&0.045M&0.045M&0.045M\\
        %     \bottomrule
        %     \end{tabular}
        %     \caption{Comparison of model sizes (\ie number of learnable parameters in millions) over different datasets. \name is 12 to 90x smaller.}
        %     \label{tab:model_size}
        %     \vspace{-5pt}
        %     \end{table}

        \begin{table*}[t]
            \centering\small
            \begin{tabular}{p{0.08\linewidth}p{\factor\linewidth}p{\factor\linewidth}p{\factor\linewidth}p{\factor\linewidth}p{\factor\linewidth}p{\factor\linewidth}
                p{\factor\linewidth}p{\factor\linewidth}p{\factor\linewidth}p{\factor\linewidth}}
            \toprule
            \textsc{S$\rightarrow$T} &\multicolumn{2}{l}{\textsc{Office}}& \multicolumn{2}{l}{\textsc{Tool}}& \multicolumn{2}{l}{\textsc{Movie}}&  \multicolumn{2}{l}{\textsc{Music}} & \multicolumn{2}{l}{\textsc{Epinions}}\\ 
            Metric  & R@10     & N@10    & R@10   & N@10   & R@10   & N@10  & R@10   & N@10 & R@10   & N@10\\
            \midrule
            \multicolumn{11}{c}{\textsc{Reference: Regular Sequential Recommendation}}\\
            MostPop  &0.450&0.272&0.459&0.274&0.586&0.361&0.519&0.327&0.438&0.296\\
            BERT4Rec~\cite{sun2019bert4rec}  &$0.541^*$&$0.358^*$&0.544&$0.350^*$&0.900&0.728&$0.816^*$&$0.602^*$&0.702&$0.512$\\
            % CL4SRec~\cite{cl4srec} &$0.550^*$&$0.358^*$&0.548&0.352&0.899&0.725&0.813&0.597&0.662&0.481\\
            \name &0.536&0.344&$0.551^*$&$0.339$&$0.908^*$&$0.738^*$&0.782&0.573&$0.795^*$&$0.580^*$\\
            \multicolumn{11}{c}{\textsc{Zero-shot Sequential Recommendation, Source$\rightarrow$Target}}\\
            Office $\rightarrow$ &---&---&\textbf{0.540}&\textbf{0.326}&0.838&0.624&0.755&0.542&0.724&0.512\\
            % Tool**** $\rightarrow$ &\textbf{0.548}&\textbf{0.336}&---&---&0.866&0.648&0.752&0.542&0.713&0.511\\
            Tool $\rightarrow$ &\textbf{0.543}&\textbf{0.332}&---&---&\textbf{0.881}&\textbf{0.659}&0.749&0.536&0.717&0.510\\
            Movie$\rightarrow$ &0.520&0.320&0.508&0.302&---&---&\textbf{0.811}&\textbf{0.600}&\textbf{0.751}&\textbf{0.537}\\
            Music$\rightarrow$ &0.503&0.310&0.496&0.312&0.836&0.636&---&---&0.739&0.518\\
            Epinions$\rightarrow$ &0.517&0.317&0.470&0.302&0.872&0.656&0.774&0.517&---&---\\
            % \midrule
            % Tool$^+$ $\rightarrow$&0.750&0.554&0.535&0.333&0.756&0.538\\
            % Movie$^+$$\rightarrow$ &0.778&0.571&0.547&0.340&0.776&0.560\\
            % Epinions$^+$$\rightarrow$ &0.787&0.578&0.527&0.318&---&---\\
            \bottomrule
            \end{tabular}
            \caption{\textit{Zero-shot recommendation results}. Results for cross-domain, cross-application zero-shot transfer. \textsc{S$\rightarrow$T} means we pre-train \name using S's data (columns) and evaluate on T's data (rows). We follow the zero-shot inference setting in~\Cref{sec:zero_shot_inference}. Reference models are trained from scratch on the target dataset. The best-performing zero-shot transfer results of each dataset are in bold. We empirically show \name achieves remarkable zero-shot generalization performance across domains.} 
            \label{tab:uu_results}
        
            \end{table*}
\subsection{Zero-shot Inference}
\label{sec:zero_shot_inference}
% Our inference setup is defined in~\Cref{sec:problem_formulation}. For zero-shot transfer, we first compute two levels of popularity dynamics on the target dataset. Then, we directly apply the model pre-trained on the source dataset to get the prediction.

Suppose we are given a pre-trained model $\mathcal{F}$ trained on $\mM$, where $\mathcal{F}$ is the scoring function learned from source domain $\mM$. Denote the interaction matrix of the target domain as $\mM'$. We first compute the popularity dynamics of each item in $\mM'$ over a coarser period and a finer period. Then, we apply the pre-trained model $\mathcal{F}$ to $\mM'$ and compute the prediction score as:

\begin{equation}
    \hat{y}(v^t_{j'}|\mathcal{S}'_u) = \mathcal{F}(v^t_{j'}|\mathcal{S}'_u, \mM')
    \label{eqn:zero_shot_inference}
\end{equation}

Note that in this procedure, we use the pre-trained model $\mathcal{F}$ trained on domain $\mM'$ to predict the next item $v^t_{j'}$ that user $u'$ will interact with in domain $\mM'$. We do not use any auxiliary information in either domain. In addition, none of the parameters in $\mathcal{F}$ are updated during the zero-shot inference process.

To summarize, in this section, we showed how to develop a pre-trained sequential recommender system based on the popularity dynamics of items. We enforce the structure of each interaction in the sequence by the positional encoding and introduce a relative time encoding for modeling time intervals between two consecutive interactions. In addition, we showed the training process and formally defined the zero-shot inference procedure. In the next section, we present experiments to evaluate \name.

%% file: experiments.tex
We present extensive experiments on five real-world datasets to evaluate the performance of \textsc{PrepRec}, following the problem settings in~\Cref{sec:problem_formulation}. We introduce the following research questions (RQ) to guide our experiments: 
\begin{enumerate*}[label=(RQ\arabic*)]
    \item How well can \name perform on zero-shot cross-domain and cross-application transfer?
    \item Why should we model popularity dynamics in sequential recommendation?
    \item What affects the performance of \name? 
    % \item How well can \name perform under the regular, in-domain sequential recommendation setting?
    % \item Can \name (popularity dynamics) generalize to zero-shot cross-domain and cross-application transfer?
    % \item How do different model components affect the performance of \name?
    % \item What is the difference between the predictions of \name and those of the state-of-the-art sequential recommenders?
    % \item What affects the performance of \name? 
\end{enumerate*}
\vspace{-10pt}
% We now present the experimental results of our proposed method. We first introduce the datasets~\cref{sec:datasets}
%  and the baselines and comparison methods~\cref{sec:zero_shot_inference}. Then, we describe the evaluation metrics~\cref{sec:evaluation_metrics}. 
%  Finally, we analyze the performance of our method~\cref{sec:performance_analysis} and conduct extensive sensitivity analyses~\cref{sec:sensitivity_analysis} 
%  and ablation studies~\cref{sec:ablation_study} to attribute this performance to various choices and components in our model. 

% \label{sec:datasets}

% The Tool dataset is the largest in terms of the number of users and items, with Office and Movie following, and Epinions and Music smallest.

%The Movie and Music
%datasets are densest, Epinions in the middle, and Office and Tool sparsest. 
\subsection{Datasets and Preprocessing}
\label{sec:datasets}    
We evaluate our proposed method on five real-world datasets across different applications, with varying sizes, and density levels. 

\textbf{Amazon}~\cite{ni2019justifying} is a series of product ratings datasets obtained from Amazon.com, split by product categories. We consider the \textit{Office} and \textit{Tool} product domains in our study. \textbf{Douban}~\cite{song2019session} consists of three datasets across different domains, collected from Douban.com, a Chinese review website. We work with the \textit{Movie} and \textit{Music} datasets. \textbf{Epinions}~\cite{tang-etal12a, tang-etal12b} is a dataset crawled from product review site Epinions. We utilize the ratings dataset for our study.
% \begin{itemize}
%     \item \textbf{Amazon}~\cite{ni2019justifying} is a series of product ratings datasets obtained from Amazon.com, split 
%     by product domain. We consider the \textbf{Office} and \textbf{Tool} product domains in our study.
%     \item \textbf{Douban}~\cite{song2019session} consists of three datasets across different domains, collected from Douban.com, 
%     a Chinese review website. We work with the \textbf{Movie} and \textbf{Music} datasets. 
%     \item \textbf{Epinions}~\cite{tang-etal12a, tang-etal12b} is a dataset crawled from product review site Epinions. We utilize the 
%     ratings dataset for our study.
% \end{itemize}

We present dataset statistics in~\Cref{tab:dataset_statistics}. We compute the density as the ratio of the number of interactions to the number of users times the number of items. Douban datasets (\ie movie and music) are the densest and have no auxiliary information available, while the Amazon review datasets (\ie office and tool) are the sparsest.

For fair evaluation, we follow the same preprocessing 
procedure as previous works~\cite{kang2018self,sun2019bert4rec},
% For all datasets, we consider all user-item interactions as implicit feedback of 1. We apply 5-core filtering to remove users and items with less than 5 interactions. 
\ie we binarize the explicit ratings to implicit feedback.
In addition, for each user, we sort interactions by their timestamp and use the second most recent action for validation, the most recent action for testing, and the rest for training. 
% As described in~\cref{sec:model_architecture}, our method directly utilizes interaction timestamps 
% to discover long-term and short-term popularity trends. 
% Note that existing sequential methods ~\cite{sun2019bert4rec,kang2018self} use timestamps for sequence construction, so our method isn't any more restrictive in application. 

    % \vspace{-0.5\baselineskip}
    \vspace{-7pt}
\subsection{Baselines and Experimental Setup}
\label{sec:experimental_setup}
\vspace{-3pt}

\noindent\textbf{Baselines:} Our baselines (supplementary materials contain detailed descriptions) include classic general recommendation models (\eg MostPop, BPR~\cite{bpr}, NCF~\cite{ncf}, LightGCN~\cite{lightgcn}) and state-of-the-art sequential recommendation models (\eg Caser~\cite{tang2018personalized}, SasRec~\cite{kang2018self}, BERT4Rec~\cite{sun2019bert4rec}, TiSasRec~\cite{li2020time}, CL4SRec~\cite{cl4srec}).

\begin{table*}[t]
    \centering
    \small
    \begin{tabular}{p{0.13\linewidth}p{\factor\linewidth}p{\factor\linewidth}p{\factor\linewidth}p{\factor\linewidth}p{\factor\linewidth}
    p{\factor\linewidth}p{\factor\linewidth}p{\factor\linewidth}p{\factor\linewidth}p{\factor\linewidth}}
    \toprule
    \textsc{Dataset} & \multicolumn{2}{l}{\textsc{Office}} & \multicolumn{2}{l}{\textsc{Tool}} & \multicolumn{2}{l}{\textsc{Movie}}& \multicolumn{2}{l}{\textsc{Music}}& \multicolumn{2}{l}{\textsc{Epinions}}\\ 
    Metric  & R@10     & N@10    & R@10   & N@10   & R@10   & N@10  & R@10   & N@10 & R@10   & N@10\\
    \midrule

    \multicolumn{11}{c}{\textsc{General Recommender Systems}}\\
    MostPop  &0.450&0.272&0.459&0.274&0.586&0.361&0.519&0.327&0.438&0.296\\
    BPR~\cite{bpr}   &0.457&0.289&0.363&0.216&0.747&0.477&0.646&0.434&0.568&0.397\\
    NCF~\cite{ncf}  &0.446&0.266&0.388&0.239&0.784&0.505&0.652&0.437&0.570&0.396\\
    % NGCF~\cite{ngcf}  &0.460&0.291&0.421&0.256&0.788&0.510&0.659&0.441&0.571&0.393\\
    LightGCN~\cite{lightgcn} &0.465&0.293&0.463&0.275&0.793&0.512&0.665&0.447&0.575&0.396\\
    \multicolumn{11}{c}{\textsc{Sequential Recommender Systems}}\\
    Caser~\cite{tang2018personalized}&0.512&0.334&0.496&0.297&0.891&0.701&0.796&0.576&0.674&0.475\\
    % MARank~\cite{yu2019multi} &0.488&0.312&0.468&0.282&0.888&0.683&0.739&0.501&0.628&0.436\\
    SasRec~\cite{kang2018self}  &0.539&0.354&0.536&0.337&0.918&0.749&0.816&0.599&$0.705^*$&0.501\\
    BERT4Rec~\cite{sun2019bert4rec}  &0.541&0.358&0.544&0.350&0.900&0.728&$0.816^*$&$0.602^*$&0.702&$0.512^*$\\
    TiSasRec~\cite{li2020time} &0.531&0.349&0.539&0.341&$0.918^*$&$0.752^*$&0.809&0.523&0.701&0.499\\
    CL4SRec~\cite{cl4srec} &$0.550^*$&$0.358^*$&$0.548^*$&$0.352^*$&0.899&0.725&0.813&0.597&0.662&0.481\\
    % DuoRec~\cite{duorec} &&&&&&&&&&\\
    \midrule
    
    \name &0.536&0.344&0.551&0.339&0.908&0.738&0.782&0.573&0.795&0.580\\
    % \name $\Delta$ &-1.5\%&-7.1\%&+0.2\%&-4.4\%&-2.5\%&-4.9\%&-4.6\%&-8.1\%&+7.3\%&+6.1\%\\
    \name $\Delta$ &-2.5\%&-3.9\%&+1.2\%&-3.6\%&-1.1\%&-1.8\%&-1.9\%&-4.8\%&+12.7\%&+13.3\%\\
    % \textsc{PrepRec}+SasRec &0.589&0.394&0.585&0.381&0.940&0.787&0.865&0.657&0.788&0.599\\
    % \textsc{Interp} &\textbf{0.648}&\textbf{0.483}&\textbf{0.652}&\textbf{0.489}&\textbf{0.940}&\textbf{0.810}&\textbf{0.901}&\textbf{0.783}&\textbf{0.821}&\textbf{0.691}\\
    \textsc{Interp} &\textbf{0.648}&\textbf{0.483}&\textbf{0.659}&\textbf{0.482}&\textbf{0.929}&\textbf{0.769}&\textbf{0.851}&\textbf{0.653}&\textbf{0.816}&\textbf{0.640}\\
    \textsc{Interp} $\Delta$ &+17.8\%&+34.9\%&+20.3\%&+35.0\%&+1.1\%&+2.3\%&+4.3\%&+8.5\%&+15.7\%&+25.0\%\\
    \bottomrule
    \end{tabular}
    \caption{\textit{Regular sequential recommendation results, RQ2, (\cref{sec:performance_analysis})}. We make bold the best results and mark the best baseline results with $'*'$. \textsc{Interp} represents the interpolation results between \name and BERT4Rec. \name $\Delta$ denotes the performance difference between \name and the best results among the selected baselines, similar for \textsc{Interp} $\Delta$. \name achieves comparable performance to the state-of-the-art sequential recommenders, with only on average 0.2\% worse than the best performing sequential recommenders in R@10 while having only a fraction of the model size (\Cref{tab:model_size}). After a simple post-hoc interpolation, we outperform the state-of-the-art sequential recommenders by 11.8\% in R@10 on average.}
    \label{tab:ss_results}
    % \vspace{-8pt}
    \end{table*}
Following previous works~\cite{koren2008factorization,ncf,kang2018self,sun2019bert4rec}, we adopt the \textit{leave-one-out} evaluation method: for each user, we pair the test item with 100 unobserved items according to the user's interaction history. Then we rank the test item for the user among the 101 total items. We use two standard evaluation metrics for top-$k$ recommendation: Recall@k (R@k) and Normalized Discounted Cumulative Gain@k (N@k). 
% R@k measures the proportion of test items that appear in the top-$k$ recommendations, and N@k accounts for the ranking position of the test interaction. 
Our model explicitly utilizes popularity information. Therefore, we also present results where we sample the negatives based on their popularities, \ie popular items have higher probabilities of being sampled as negatives. We report the average of R@k and N@k over all the test interactions.

We use publicly available implementations for the baselines. For fair evaluation, we set dimension size $d$ to 50, max sequence length $L$ to $200$, and batch size to $128$ in all models. We use an Adam optimizer and tune the learning rate in the range $\{10^{-4}, 10^{-3}, 10^{-2}\}$ and set the weight decay to $10^{-5}$. We use the dropout regularization rate of $0.3$ for all models. We set $\gamma = 0.5$ in \Cref{eqn:popularity_gamma}, whose reason we will discuss the reason in supplement materials. We define the coarse and fine period to be 10 and 2 days respectively, and we fix the window size to be $m = 12$ and $n =4$ for all datasets (\Cref{sec:item_trajectory_encoder}). We train \name for a maximum of 80 epochs. All experiments are conducted on a Tesla V100 using PyTorch. We repeat each experiment 5 times with different random seeds and report the average performance.
%  For our model, we use existing implementation hyperparameters with minimal tuning, training through Adam with learning rate 0.001, $\beta_1 = 0.9,$ and $\beta_2 = 0.98$. We choose annual calendar
% months and weeks as the long-term and short-term time horizons, respectively, with 12-month and 4-week windows for each item feature vector at any time.
% We set $\gamma = 0.5$ for percentile calculation, use sinusoidal positional encoding, and use sinusoidal relative time encoding. These decisions are studied further in~\cref{sec:sensitivity_analysis} and~\cref{sec:ablation_study}.
% \vspace{-13pt}
\begin{figure}[b]
    \centering
    \includegraphics[width=0.8\linewidth]{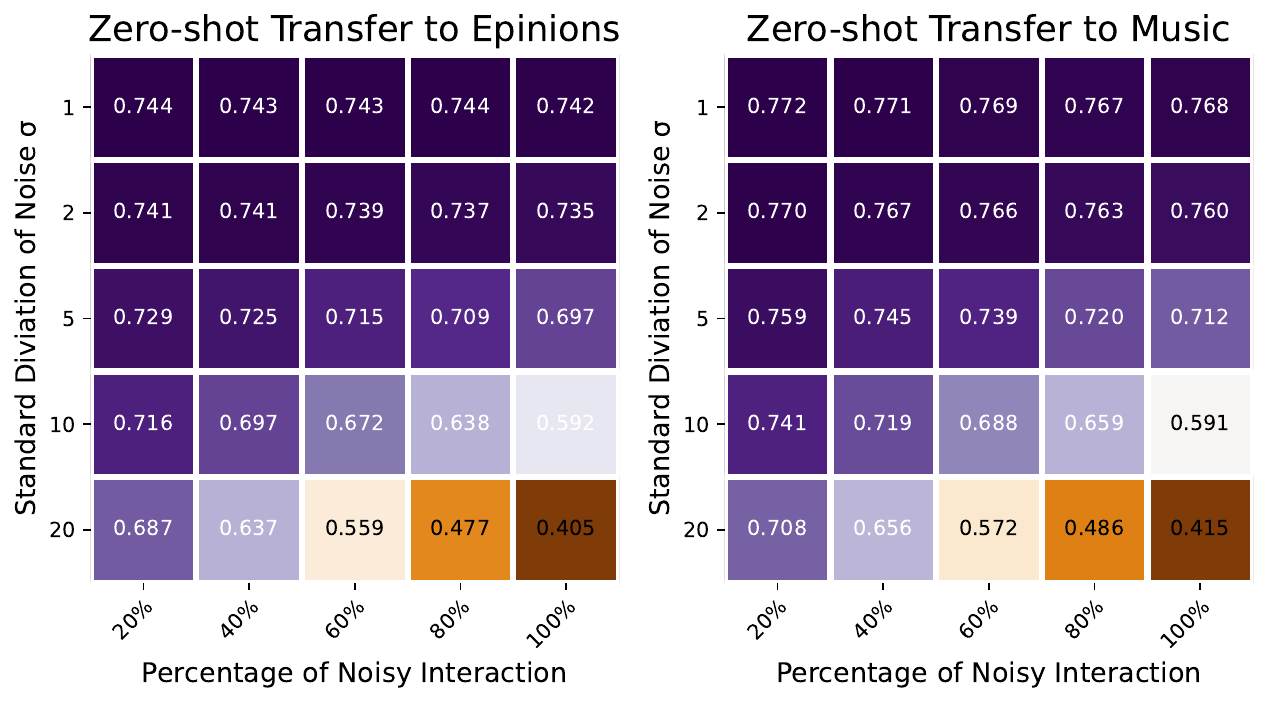}
    \caption{Zero-shot Transfer Results (R@10) with Gaussian noise added to the item popularity statistics (\cref{sec:robustness_to_noise}). We find that \name is relatively robust to noise.}
    \label{fig:noise}
    \vspace{-5pt}
\end{figure}
% \vspace{-\baselineskip} 
\vspace{-10pt}
    \subsection{Zero-shot Transferability (RQ1)}
    % \vspace{-3pt}
    \label{sec:zero_few_shot}
    \subsubsection{Zero-shot Transfer Results}
    We follow the zero-shot inference setting introduced in~\Cref{sec:zero_shot_inference} and report the results in~\Cref{tab:uu_results}. We also include the results of \name and the best-performing sequential recommenders trained on the target dataset for reference. In the zero-shot setting, \name shows minimal performance reduction in the target datasets (\ie 6\% maximum and 2\% average reduction in R@10). The best zero-shot transfer results from \name only fall short against the selected sequential recommendation baselines by up to 4\% and even outperform (by up to 6.5\%) them on the Epinions and Office. We found that \name trained on douban-movie and amazon-tools show the highest generalizability, even outperforming the target-trained models on Music (0.811 \textit{vs.} 0.782 on R@10). We conjecture that this is because Movie is the largest dataset in terms of the number of interactions. Overall, these results show \name's effectiveness in zero-shot transfer without any training on interaction data or side information. In addition, this experiment also demonstrates that the popularity dynamics-based item and sequence representations are generalizable across domains.

    \subsubsection{Robustness to Noise}
    \label{sec:robustness_to_noise}

        % \vspace{-8pt}
    % \vspace{-4pt}

    We further investigate the robustness of \name to possible noise in zero-shot transfer by adding Gaussian noise $ \sim \mathcal{N}(0, \sigma)$ to the item popularity statistics and evaluate the zero-shot transfer performance on Douban-Music and Epinions from Douban-Movie. We randomly choose some percentage of items in the sequence to add noise, as indicated in ~\Cref{fig:noise}. We find that \name is relatively robust to noise, maintaining robust performance across different noise levels at 20\% noised interaction. We attribute this to the model's ability to learn from the overall popularity dynamics, which is less affected by noise in individual item popularity statistics. In addition, when the noise level is relatively low, \eg $\sigma \leq 5$, even if 100\% of the sequence is noised, \name still holds the performance, indicating significant item popularity shifts exist in the sequence (\Cref{fig:intro}).
    
    \begin{table}[h]
        \centering
            \small
        \begin{tabular}{p{0.22\linewidth}p{0.08\linewidth}p{0.08\linewidth}p{0.08\linewidth}p{0.08\linewidth}p{0.08\linewidth}
        p{0.08\linewidth}}
        \toprule
        \textsc{Dataset} & \multicolumn{2}{l}{\textsc{Music}} & \multicolumn{2}{l}{\textsc{Office}}& \multicolumn{2}{l}{\textsc{Epinions}}\\ 
        Metric  & R@10     & N@10    & R@10   & N@10   & R@10   & N@10  \\
    
        % \textsc{Interp} &\textbf{0.871}&\textbf{0.745}&\textbf{0.613}&\textbf{0.469}&\textbf{0.824}&\textbf{0.690}\\
        \midrule
        MostPop &0.197&0.139&0.099&0.046&0.163&0.110\\
        SasRec~\cite{kang2018self} &\textbf{0.749}&0.519&0.453&0.291&0.658&0.442\\
        BERT4Rec~\cite{sun2019bert4rec} &0.747&0.519&\textbf{0.461}&\textbf{0.299}&0.655&0.456\\
        \midrule
        \name &0.739&\textbf{0.523}&0.443&0.280&\textbf{0.762}&\textbf{0.551}\\
        \name $\Delta$ &-1.3\%&+0.7\%&-2.2\%&-6.3\%&+15.8\%&+17.2\%\\
        % \midrule
        % Sas+BERT &0.853&0.738&0.584&0.450&0.765&0.641\\
    
        % t-3 &0.780&0.543&0.539&0.337&0.723&0.514\\
        % t-6 &0.730&0.497&0.523&0.326&0.682&0.473\\
        % t-12 &0.705&0.462&0.507&0.313&0.645&0.445\\
        \bottomrule
        \end{tabular}
        \caption{\textit{Regular sequential recommendation results (\cref{sec:performance_analysis}) with popularity-based negative sampling}. \name can learn discriminative item and sequence representations despite depending only on popularity statistics.}
        \label{tab:popularity_based_negative_sampling}
        \vspace{-5pt}
        \end{table}
% \vspace{-\baselineskip}
\vspace{-10pt}
\subsection{Why Popularity Dynamics? (RQ2)}
\vspace{-3pt}
\name shows excellent performance in zero-shot sequential recommendation. Therefore, we ask: what is the role of popularity dynamics in sequential recommendation, and how much does it explain the performance of state-of-the-art sequential recommenders? Therefore, we propose the following experiments to investigate the importance of popularity dynamics in sequential recommendation.
% \vspace{-10pt}

 \subsubsection{Regular Sequential Recommendation (RQ2)}
% \subsection{Experimental Results}
\label{sec:performance_analysis}

We show comparisons of \name against state-of-the-art sequential recommenders in the regular sequential recommendation tasks (\Cref{tab:ss_results}), \ie all models are trained from scratch. \name achieves competitive performance---within 2\% in R@10 and 5\% in N@10, with the state-of-the-art baselines. On Epinions, \name even outperforms all baselines by 7.3\%,  particularly impressive since \name has significantly fewer model parameters (\Cref{tab:model_size}). 

\begin{table}[t]
    \centering
        \small
    \begin{tabular}{p{0.12\linewidth}p{0.1\linewidth}p{0.1\linewidth}p{0.1\linewidth}p{0.1\linewidth}p{0.1\linewidth}
    p{0.1\linewidth}}
    \toprule
    \textsc{Dataset}& {\textsc{Office}} & {\textsc{Tool}} & {\textsc{Movie}} & {\textsc{Music}} & {\textsc{Epinions}}\\ 
    \midrule
    SasRec &1.331M&3.581M&2.044M&0.542M&1.054M\\
    BERT4Rec &2.687M&7.233M&4.126M&1.094M&2.127M\\
    TiSasRec &1.367M&3.617M&2.127M&0.578M&1.09M\\
    \name &0.045M&0.045M&0.045M&0.045M&0.045M\\
    \bottomrule
    \end{tabular}
    \caption{Comparison of model sizes (\ie number of learnable parameters in millions) over different datasets. \name is 12 to 90x smaller.}
    \label{tab:model_size}
    \vspace{-5pt}
    \end{table}

\name explicitly models popularity information and the MostPop demonstrates decent performance compared to the remaining baselines, thus we conduct an additional experiment (\Cref{tab:popularity_based_negative_sampling}) where we sample the unobserved (negative) items based on their popularity~\cite{sun2019bert4rec}. As shown in~\Cref{tab:popularity_based_negative_sampling}, MostPop's performance dropped significantly, while \name shows even more competitive performance on some datasets (\eg Music and Epinions). This suggests \name learns discriminative item and sequence representations.

\name learns item representations through popularity dynamics, which is conceptually different from learning representations specific to each item ID. Therefore, we propose a simple post-hoc interpolation to investigate how much can popularity dynamics explain the performance of state-of-the-art sequential recommenders. We interpolate the scores from \name with the scores from BERT4Rec as follows: $\hat{y}_{intp}(v^t_j|\mathcal{S}_u) = \alpha*\hat{y}_{O}(v^t_j|\mathcal{S}_u) + (1-\alpha)*\hat{y}_S(v_j|\mathcal{S}_u)$, where $\hat{y}_{O}(v^t_j|\mathcal{S}_u)$ and $\hat{y}_S(v_j|\mathcal{S}_u)$ are the scores from \name(\Cref{eqn:prediction}) and BERT4Rec, respectively. We set $\alpha = 0.5$ for all datasets.  After interpolation, the performance significantly boosts by up to 34.9\% in N@10. Gains are largest in the medium and low-density datasets (Epinions, Amazon), indicating that our model complements existing methods in sparse datasets where item embeddings are less informative. Therefore, it is crucial to consider popularity dynamics to maximize performance.
% \vspace{-7pt}
\subsubsection{Qualitative Analysis on Regular Sequential Recommendation}
\label{sec:qualitative_analysis}
\begin{figure}[h]
    \centering
    \includegraphics[width=\linewidth]{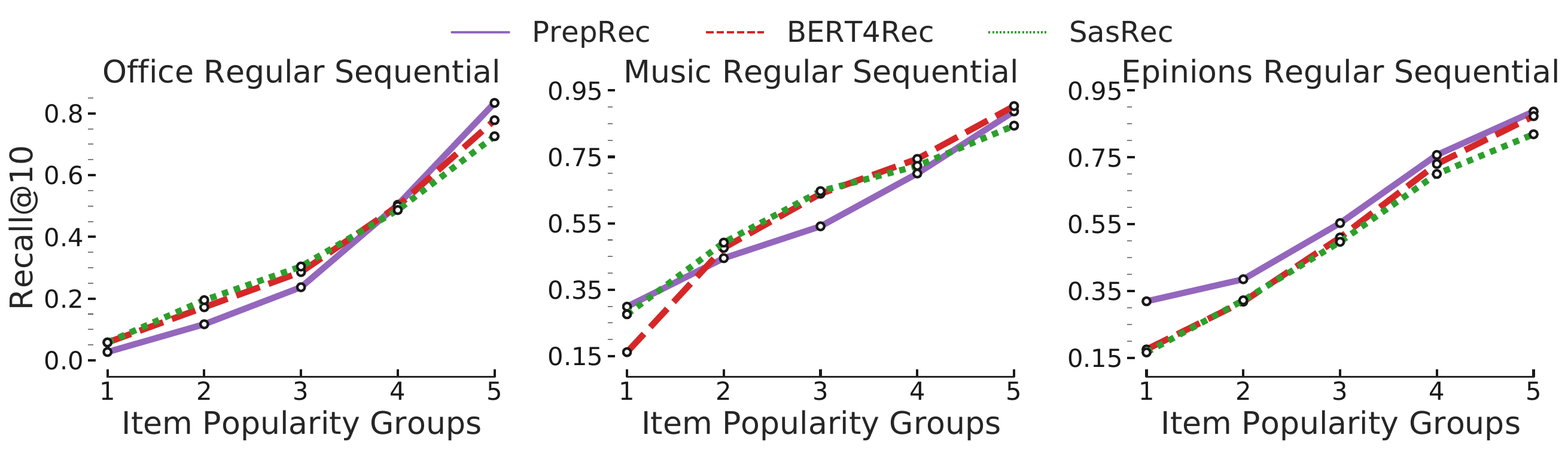}
    \caption{Recommendation results for different item popularity groups (\cref{sec:qualitative_analysis}), where Group 1 represents the least popular items, and Group 5 represents the most popular items. \name achieves better performance on long-tail items while having competitive performance on popular items.}
    \label{fig:item_quality}
    % \vspace{-5pt}
\end{figure}
% \subsection{Effect of Training Data Available(RQ4)}
% \label{sec:sensitivity_analysis}
We analyze the performance of \name in detail. We separate test items into equally sized groups based on their popularity in the training set then compute the average R@10 and N@10 for each group (\Cref{fig:item_quality}). \name achieves better performance on item group with the least interactions, \ie long-tail items, while the SasRec and BERT4Rec show stronger performance on popular items. Long-tail item recommendation is a particularly challenging task explored by many previous works~\cite{protocf} and requires recommenders able to learn high-quality representations with just a few interactions. This corresponds to our observation that \name is more robust to data sparsity and can learn discriminative item and sequence representations (\Cref{sec:performance_analysis}), showing that long-tail item recommendation can benefit from \name's popularity dynamics-based item representations.
\renewcommand*{\factor}{0.05}
% \vspace{-0.75\baselineskip}
\vspace{-7.5pt}

\subsection{What affects \name performance? (RQ3)}
\label{sec:rq3}
\subsubsection{Ablation Study}
% \vspace{-5pt}
\label{sec:ablation_study}
\begin{table}[h]
    \centering
    \small
    \begin{tabular}{p{0.40\linewidth}p{\factor\linewidth}p{\factor\linewidth}p{\factor\linewidth}p{\factor\linewidth}p{\factor\linewidth}
    p{\factor\linewidth}}
    \toprule
    \textsc{Dataset} & \multicolumn{2}{l}{\textsc{Music}} & \multicolumn{2}{l}{\textsc{Office}}& \multicolumn{2}{l}{\textsc{Epinions}}\\ 
    Metric  & R@10     & N@10    & R@10   & N@10   & R@10   & N@10  \\
    \midrule
    \name  &0.782&0.573&0.536&\textbf{0.344}&\textbf{0.795}&\textbf{0.580}\\
    w/o Relative Time $\mT$ (\Cref{sec:time_interval_encoder}) &0.734&0.514&\textbf{0.541}&0.334&0.782&0.562\\
    w/o Positional $\mP$ (\Cref{sec:positional_encoding})&0.765&0.544&0.530&0.332&0.772&0.554\\

    w/o Popularity Dynamics $\mathcal{P}$ &\textbf{0.800}&\textbf{0.594}&0.530&0.341&0.761&0.560\\
    w/o Popularity Dynamics $\mathcal{H}$ &0.705&0.582&0.525&0.337&0.730&0.533\\

    Sinuoisal Popularity Encoding&0.779&0.570&0.529&0.340&0.772&0.561\\

    \bottomrule
    \end{tabular}
    \caption{Ablation study of \name's different variants. }
    \label{tab:Ablation Study}
    \vspace{-3pt}
    \end{table}

    % \vspace{-2pt}
    % \subsubsection{Time and Positional Encoding Ablations}
    Here, we assess the importance of different components crucial to \name, \ie relative time encoding (\Cref{sec:time_interval_encoder}), positional encoding (\Cref{sec:positional_encoding}), popularity encoder $E_p$ (\Cref{sec:item_popularity_encoding}), and resolutions for popularity dynamics (\Cref{sec:item_trajectory_encoder}). We find that removing relative time encoding $\mT$ results in the largest performance drop on both the Music and Office datasets. This suggests that the relative time encoding is crucial for effectively capturing the popularity dynamics. Removing positional encoding $\mP$ results in a maximum of 2.2\% drop in R@10 on the Office dataset, indicating positional encoding is important for capturing sequential information. In addition, changing $E_p$ to the non-linear sunusoidal encoding shows worse performance on all datasets, meaning that the linear encoding is more suitable for capturing the popularity dynamics. 
    On the music dataset, removing coarse popularity encoding $\mathcal{P}$ results improves the performance by 2\% in R@10, while removing fine popularity encoding $\mathcal{H}$ results in a 7.5\% drop in R@10. This suggests that the music domain is more sensitive to recent trends in popularity. Coarse and fine popularity encodings complement each other on other datasets.

        \begin{table}[h]
            \centering
            \small
            \begin{tabular}{p{0.40\linewidth}p{\factor\linewidth}p{\factor\linewidth}p{\factor\linewidth}p{\factor\linewidth}p{\factor\linewidth}
            p{\factor\linewidth}}
            \toprule
            \textsc{Dataset} & \multicolumn{2}{l}{\textsc{Music}} & \multicolumn{2}{l}{\textsc{Office}}& \multicolumn{2}{l}{\textsc{Epinions}}\\ 
            Metric  & R@10     & N@10    & R@10   & N@10   & R@10   & N@10  \\
            \midrule
            $\gamma = 0$ (\textsc{Curr-pop}) &0.749&0.542&0.512&0.328&0.689&0.496\\
            $\gamma = 0.25 $ &0.764 & 0.529&0.538&0.338&0.761&0.562\\
            $\gamma = 0.5^*$ (\textsc{weighted -pop}) &\textbf{0.782}&\textbf{0.573}&\textbf{0.536}&\textbf{0.344}&\textbf{0.795}&\textbf{0.580}\\
            $\gamma = 0.75$ &0.755&0.520&0.543&0.336&0.747&0.519\\
            
            $\gamma = 1$ (\textsc{cumul-pop}) &0.695&0.452&0.530&0.330&0.733&0.505\\
            \bottomrule
            \end{tabular}
            \caption{Recommendation results for varying the discounting factor $\gamma$ in~\Cref{sec:item_trajectory_encoder}. $\gamma = 0.5$ is the default setting, denoted by $'*'$. We find that $\gamma = 0.5$ generally outperforms the other two settings}
            \label{tab:gamma}
            \vspace{-5pt}
            \end{table}

\subsubsection{Effect of discounting factor $\gamma$}
\label{sec:gamma}

We examine the effect of different preprocessing weights $\gamma$ used in popularity calculation (\Cref{sec:item_popularity_encoding}). In particular, $\gamma = 1$ corresponds to the cumulative popularity, or in other words, at a given time period $t$, the overall number of interactions up to period $t$. On the other hand, $\gamma = 0$ corresponds to the current popularity, or percentiles are calculated over interactions just in $t$, same as $b_j^t$ in \Cref{eqn:popularity_gamma}.  When $\gamma = 0.5$, it can be interpreted as interactions being exponentially weighted by time, with a half-life of 1 time period.
% We compare our $\gamma = 0.5$, denoted \textsc{weight-pop}, with 
%$\gamma = 0$, denoted 
%\textsc{curr-pop}, and $\gamma = 1$, denoted \textsc{cumul-pop}. 
We find that $\gamma = 0.5$ outperforms the other two settings, with the largest gains of around 12\% R@10 and 27\% N@10 over \textsc{cumul-pop} in the dense Music dataset, 
and the largest gains over \textsc{curr-pop} in the sparser Office (5\% N@10 and 4\% N@10) and Epinions ($15\%$ R@10 and 17\%N@10) datasets. We suspect 
this is due to cumulative measures in denser 
datasets failing to capture recent trends due to the large historical presence, while current-only measures in sparser datasets 
convey too little or noisy information and lose the information of long-term trends. \textsc{curr-pop} shows decent performance on the Music dataset, suggesting that Music trends might be more cyclical and thus the current popularity is more informative.
\subsubsection{Effect of Different Time Horizons}
    \label{sec:time_horizon}
    \renewcommand*{\factor}{0.05}

    \begin{table}[h]
        \centering
        \small
        \begin{tabular}{p{0.40\linewidth}p{\factor\linewidth}p{\factor\linewidth}p{\factor\linewidth}p{\factor\linewidth}p{\factor\linewidth}
        p{\factor\linewidth}}
        \toprule
        \textsc{Dataset} & \multicolumn{2}{l}{\textsc{Music}} & \multicolumn{2}{l}{\textsc{Office}}& \multicolumn{2}{l}{\textsc{Epinions}}\\ 
        Metric  & R@10     & N@10    & R@10   & N@10   & R@10   & N@10  \\
        \midrule
        Fine:2 days; Coarse:10 days $^*$&\textbf{0.782}&\textbf{0.573}&0.536&\textbf{0.344}&\textbf{0.795}&\textbf{0.580}\\
        Fine:4 days; Coarse:15 days &0.778&0.553&\textbf{0.537}&0.341&0.790&0.574\\
        Fine:7 days; Coarse:30 days &0.760&0.509&0.526&0.334&0.757&0.543\\
        \bottomrule
        \end{tabular}
        \caption{Recommendation results for varying time horizons. Fine and coarse time horizons are used for short-term and long-term popularity dynamics respectively (\Cref{sec:item_popularity_encoding}). }
        \label{tab:gamma}
        \vspace{-5pt}
        \end{table}
    We study the effect of different time horizons to \name. We found that in general, long-term horizons of 30 days and short-term horizons of 7 days perform worse than the other settings. This is likely because the long-term horizon might lead to the lack of resolutions in popularity statistics. We also find that depending on the dataset, the effect of different time horizons also varies. For example, both Music and Epinions show larger performance decrease from short to long-term horizons than Office. This could be because Music and Epinions are more sensitive to recent trends than Office, or their data are denser in terms of time granularity.

\vspace{-10pt}
\subsection{Fine-tune Capability}
% \vspace{-3pt}
\label{sec:fine_tune}
\begin{table}[h]
    \centering
    \small
    \begin{tabular}{p{0.20\linewidth}p{\factor\linewidth}p{\factor\linewidth}p{\factor\linewidth}p{\factor\linewidth}p{\factor\linewidth}
    p{\factor\linewidth}}
    \toprule
    \textsc{Dataset} & \multicolumn{2}{l}{\textsc{Movie}$\rightarrow$\textsc{Music}} & \multicolumn{2}{l}{\textsc{Tool}$\rightarrow$\textsc{Office}}& \multicolumn{2}{l}{\textsc{Tool}$\rightarrow$\textsc{Epinions}}\\ 
    Metric  & R@10     & N@10    & R@10   & N@10   & R@10   & N@10  \\
    \midrule
    \name &0.803&0.591&\textbf{0.472}&\textbf{0.300}&\textbf{0.489}&\textbf{0.264}\\
    SasRec &0.815&0.599&0.437&0.290&0.433&0.245\\
    BERT4Rec &\textbf{0.816}&\textbf{0.602}&0.407&0.249&0.433&0.255\\
    \bottomrule
    \end{tabular}
    \caption{Recommendation results for fine-tuning \textsc{PrepRec}. We fine-tune \name and retrain the baselines from scratch on the target dataset.}
    \label{tab:fine_tune}
    % \vspace{-2pt}
    \end{table}
    % \vspace{-0.5\baselineskip}
We also investigate \textsc{PrepRec}'s fine-tune capability. To ensure target datasets are smaller than the source, we further process the target datasets such that they are no more than 10\% of the source datasets' total interactions. After further processing, we follow the same experimental setup in~\Cref{sec:experimental_setup}. We fine-tune \name and retrain the baselines from scratch on the target dataset and report the results in~\Cref{tab:fine_tune}.  We find that \name, after fine-tuning, outperforms the selected baselines on Office and Epinions by up to 12.9\%, indicating that \name is capable of learning from the limited data and can be further fine-tuned to achieve better performance.
% \vspace{-\baselineskip}
% \vspace{5pt}
\vspace{-8pt}

\subsection{Discussion}
\label{sec:discussion}
\vspace{-2pt}
\name demonstrates the strong ability for zero-shot transfer.
%  In addition, its prediction captures the popularity shifts in the sequence and is complementary to state-of-the-art sequential recommenders.
We argue that \name is particularly useful in the following scenarios: 1) initial sequential model when the data in the domain is sparse; 2) backbone for developing more complex sequential recommenders (\ie prediction interpolation) 3) online recommendation settings.

\name captures the popularity shifts in the sequence and is complementary to state-of-the-art sequential recommenders. It is worth noting that item popularity dynamics might not capture everything in users' preferences, but we believe they are orthogonal components towards capturing user preferences, which could explain why the interpolation results substantially outperform both \name and the selected state-of-the-art baselines (\Cref{tab:ss_results}). 
 
Additionally, time granularity is also crucial for popularity dynamics, and sequence analysis requires careful consideration of the time horizon. Intuitively, when the dataset time precision is less accurate, i.e., weeks or days, we expect the performance to decrease as the sequential information and popularity dynamics become muddled. If the time precision in the training data increases, we can expect more accurate user sequences and more accurate measures of popularity dynamics. In general, time precision will not significantly impact the performance of \name in most scenarios as in practice, online platforms can record precise time data for each user-item interaction. We will include more discussion in the arXiv version of this paper~\cite{preprec}.

%% file: conclusion.tex
\vspace{-2pt}

% \name demonstrates the ability for zero-shot transfer. In addition, its prediction captures the popularity shifts in the sequence and is complementary to state-of-the-art sequential recommenders. 

% We propose \textsc{PrepRec}, a pre-trained sequential recommendation model capable of zero-shot cross-domain transfer. 
%We assume no access to auxiliary information (\eg textual data), and we set a baseline for pre-trained sequential recommenders under the most challenging setting.

% In this paper, we developed universal, transferable item representations for the zero-shot, cross-domain setting based on the popularity dynamics of items, without any auxiliary information. 

In this paper, using the critical insight of popularity dynamics in the user's sequence, we developed a novel pre-trained sequential recommendation framework, \textsc{PrepRec}, for the zero-shot, cross-domain setting without any auxiliary information. \name learned transferable, universal item representations via popularity dynamics-aware transformers. We empirically showed that \name can achieve excellent zero-shot transfer to a target domain, comparable to state-of-the-art sequential recommenders trained on the target domain. With extensive within-domain experiments, we found performance gains of 11.8\% when we interpolated \name's results with state-of-the-art sequential recommenders, indicating that \name is learning complementary information. We posit that popularity dynamics are crucial for developing generalizable sequential recommenders.

As part of future work, we plan to investigate: 1) developing more complex sequential recommenders by using \name as a backbone (\ie prediction interpolation and auxiliary information), and 2) exploring online recommendation settings.